\documentclass[12pt]{oxarticle}
\pdfoutput=1

\usepackage{amsmath, amsthm, amsfonts, amssymb, graphicx, array, float, xspace, bbm}
\usepackage[utf8]{inputenc}
\usepackage[mathscr]{eucal}
\usepackage{color}
\usepackage[table,svgnames]{xcolor}
\usepackage{booktabs}
\usepackage{slashbox}
\usepackage[bf]{caption}
\usepackage{multirow}

\usepackage{ifthen}
\newboolean{usehyperref}
\setboolean{usehyperref}{false}
\ifthenelse{\boolean{usehyperref}}{
  \usepackage[backref,linktocpage,bookmarks,plainpages=False]{hyperref}
  \usepackage{hypbmsec}
  \newcommand{\fref}[1]{\autoref{#1}}
  \newcommand{\tref}[1]{\autoref{#1}}
  \newcommand{\sref}[1]{\autoref{#1}}
  \usepackage{twoopt}
}{
  \newcommand{\fref}[1]{Figure~\ref{#1}}
  \newcommand{\tref}[1]{Table~\ref{#1}}
  \newcommand{\sref}[1]{Section~\ref{#1}}
  \newcommand{\sectionhyp}[1][]{\sectionhyprelay}
  \newcommand{\sectionhyprelay}[1][]{\sectionhyprelayrelay}
  \newcommand{\sectionhyprelayrelay}[1]{\section{#1}}
  \newcommand{\subsectionhyp}[1][]{\subsectionhyprelay}
  \newcommand{\subsectionhyprelay}[1][]{\subsectionhyprelayrelay}
  \newcommand{\subsectionhyprelayrelay}[1]{\subsection{#1}}
}
\usepackage[sort&compress, comma, square, numbers]{natbib}

\let\SS=\S 

\renewcommand{\a}{\alpha}
\renewcommand{\b}{\beta}
\newcommand{\g}{\gamma}\newcommand{\G}{\Gamma}
\renewcommand{\d}{\delta}\newcommand{\D}{\Delta}

\newcommand{\z}{\zeta}
\newcommand{\vth}{\vartheta}

\newcommand{\p}{\pi}\renewcommand{\P}{\Pi}\newcommand{\vp}{\varpi}
\renewcommand{\r}{\rho}
\newcommand{\s}{\sigma}\renewcommand{\S}{\Sigma}

\newcommand{\vph}{\varphi}

\DeclareFontFamily{OT1}{pzc}{}
\DeclareFontShape{OT1}{pzc}{m}{it}{<-> s * [1.200] pzcmi7t}{}
\DeclareMathAlphabet{\mathpzc}{OT1}{pzc}{m}{it}

\newcommand{\ccF}{\mathpzc F}

\newcommand{\ccK}{\mathpzc K}
\newcommand{\cL}{\mathcal{L}}
\newcommand{\cM}{\mathcal{M}}\newcommand{\ccM}{\mathpzc M}

\newcommand{\cO}{\mathcal{O}}
\newcommand{\cP}{\mathcal{P}}

\newcommand{\ccS}{\mathpzc S}

\newcommand{\cW}{\mathcal{W}}

\newcommand{\IK}{\mathbb{K}}

\newcommand{\IN}{\mathbb{N}}

\newcommand{\IP}{\mathbb{P}}
\newcommand{\IQ}{\mathbb{Q}}
\newcommand{\IR}{\mathbb{R}}

\newcommand{\IZ}{\mathbb{Z}}

\newcommand{\one}{\mathbbm{1}}

\font\twentyfourrm=cmr12 at 24pt
\font\eightrm=cmr8 at 8pt

\font\csc=cmcsc10
\newcommand{\beq}{\begin{equation}}
\newcommand{\eeq}{\end{equation}}
\newcommand{\beqnn}{\begin{equation*}}
\newcommand{\eeqnn}{\end{equation*}}

\newcommand{\comment}[1]{}
\newcommand{\pd}[2]{\frac{\partial #1}{\partial #2}}

\newcommand{\+}{\phantom{-}}

\font\frak eufm10 at 12pt

\font\ninefrak eufm10 at 9pt
\newcommand{\euf}[1]{\text{\frak #1}}
\newcommand{\goth}{\euf{G}}

\newcommand{\smallgoth}{\text{\ninefrak G}}
\newcommand{\YG}{Y_{\hskip-1pt\smallgoth}}

\newcommand{\Ysplit}{\kern-2pt\hbox{\Bigcheck\kern-11pt{$Y$}}}

\newcommand{\cy}{Calabi--Yau\xspace}
\newcommand{\cym}{Calabi--Yau manifold\xspace}
\newcommand{\cys}{Calabi--Yau manifolds\xspace}

\newcommand{\quotient}[1]{_{\hskip-2pt\lower1pt\hbox{$/$}\lower2pt\hbox{\hskip-1pt$#1$}}}
\newcommand{\symm}[1]{_{\hskip-3pt\lower3pt\hbox{$\left\{#1\right\}$}}}

\newcommand{\cicystop}{~\lower8pt\hbox{.}}
\newcommand{\Bigcheck}{\lower3.8pt\hbox{\smash{\hbox{{\twentyfourrm \v{}}}}}}
\newcommand{\Xcheck}{\kern0pt\hbox{\Bigcheck\kern-12.5pt{$X$}}}
\newcommand{\cicy}[2]{\begin{matrix} #1\end{matrix}\!\left[\begin{matrix}#2 \end{matrix}\right]}

\newcommand{\ii}{\text{i}}
\newcommand{\ee}{\text{e}}
\newcommand{\dd}{\text{d}}

\newcommand{\Sp}{\text{Sp}}

\newcommand{\smallfrac}[2]{\frac{\scriptstyle #1}{\scriptstyle #2}}
\newcommand{\capt}[3]{\parbox{#1}{\renewcommand{\baselinestretch}{1.0}
                                                           \caption{\label{#2}\small\it #3}}}

\renewcommand{\capt}[3]{\caption{#3}\label{#2}}

\def\place#1#2#3{\vbox to0pt{\kern-\parskip\kern-7pt
                             \kern-#2truein\hbox{\kern#1truein #3}
                             \vss}\nointerlineskip}

\InputIfFileExists{Debug}{}{}

\newcommand{\eqdef}{%
  \mathrel{\lower.1mm
    \hbox{$\stackrel{\lower.424ex\hbox{\scriptsize def}}{=}$}}
}

\newcommand{\Q}{\ensuremath{{\mathbb{Q}}}}
\newcommand{\C}{\ensuremath{{\mathbb{C}}}}
\newcommand{\Z}{\mathbb{Z}}
\newcommand{\CP}{{\ensuremath{\mathop{\null {\mathbb{P}}}\nolimits}}}

\newenvironment{smallarray}[1]
 {\null\,\vcenter\bgroup\scriptsize
  \renewcommand{\arraystretch}{0.7}%
  \arraycolsep=.13885em
  \hbox\bgroup$\array{@{}#1@{}}}
 {\endarray$\egroup\egroup\,\null}

\DeclareMathOperator{\Span}{span}
\DeclareMathOperator{\re}{re}
\DeclareMathOperator{\im}{im}

\newcommand{\Xt}{{\ensuremath{\widetilde{X}}}}

\newcommand{\dP}{\text{dP}}
\newcommand{\Fprepotential}{\ccF}
\newcommand{\Fpert}{\ccF^\text{pert}}
\newcommand{\Fnonpert}{\ccF^\text{np}}
\newcommand{\Kcone}{\ccK}

\newcommand{\url}[1]{{\tt #1}}
\begin{document}
\renewcommand{\baselinestretch}{1.1}
\setlength{\doublerulesep}{3pt}
\numberwithin{equation}{section}
\proofmodefalse
\pagestyle{empty}
\vspace*{1in}
\begin{center}
{\Huge Two One-Parameter Special Geometries}\\[2cm]
{\csc Volker Braun${}^1$,  Philip Candelas${}^2$\\ and\\ Xenia de la Ossa${}^2$}\\[2cm]
{\it ${}^1$Elsenstrasse 35\hphantom{${}^1$}\\
12435 Berlin\\
Germany\\[0.7cm]
${}^2$Mathematical Institute\hphantom{${}^2$}\\
University of Oxford\\
Radcliffe Observatory Quarter\\ 
Woodstock Road\\ 
Oxford OX2 6GG, UK}
\vfill
{\bf Abstract\\[2ex]}
\parbox{6.0in}{\setlength{\baselineskip}{14pt}
The special geometries of two recently discovered Calabi-Yau threefolds with $h^{11}{\,=\,}1$ are analyzed in detail. These correspond to the 'minimal three-generation' manifolds with $h^{21}{\,=\,}4$ and the `24-cell' threefolds with
$h^{21}=1$. It turns out that the one-dimensional complex structure moduli spaces for these manifolds are both very similar and surprisingly complicated. Both have 6 hyperconifold points and, in addition, there are singularities of the Picard-Fuchs equation where the threefold is smooth but the Yukawa coupling vanishes. Their fundamental periods are the generating functions of lattice walks, and we use this fact to explain why the singularities are all at real values of the complex structure.}
\end{center}
\newpage
\tableofcontents
\newpage
\setcounter{page}{1}
\pagestyle{plain}
\section{Introduction}
Detailed descriptions of special geometries are known only in relatively few cases. Explicit examples have been limited to dimension $\leq 3$, with the most explicit examples of 1 and 2 dimensions. In the
Kreuzer-Skarke list~\cite{Kreuzer:2000xy, Kreuzer:2004lp} there are 5
manifolds with $h^{21}=1$. All of these share features with the
parameter space of the mirror quintic in that the parameter space is a
$\IP^1$ with three singular points: a large complex structure limit,
with maximal unipotent monodromy, a Gepner point, with $\IZ_n$
monodromy, for some $n$, and a conifold point. Doran and
Morgan~\cite{Doran:2005mw} have classified the $14$ possible one-parameter
special geometries with three singular points of the these types. Rodriguez-Villegas arrived at this same 
list~\cite{Rodriguez:2001} by means of number-theoretic considerations. We now
know~\cite{Batyrev:1993vs, Font:1992uk, Clingher:2013an} Calabi-Yau manifolds corresponding to
each of these.

The two manifolds whose parameter spaces we examine here are a manifold with Hodge numbers 
$(h^{11},h^{21})=(1,4)$, which is realised as the quotient of a hypersurface in $dP_6{\times}dP_6$ by a freely acting group $G$ of order 12. There are, in reality, two variants of the is first `manifold' since there are two choices for $G$, which is either $\IZ_{12}$ or the nonabelian group $\text{Dyc}_3$.
This space can be realised as the CICY
\beqnn
\cicy{\IP^2\\ \IP^2\\ \IP^2\\ \IP^2\\}{1& 1& 1& 0& 0\\
                                                          0& 0& 1& 1& 1\\
                                                          1& 1& 1& 0& 0\\
                                                          0& 0& 1& 1& 1\\}^{(1,4)}\hskip-22pt\raisebox{-30pt}{/G}
\eeqnn
or as a reflexive polyhedron, corresponding to the fact that the
covering space is a hypersurface in the toric variety
$\dP_6{\times}\dP_6$. A detailed description can be found in
\cite{Braun:2009qy, Candelas:2008wb}.

The other manifold has $(h^{11},h^{21})=(1,1)$ and occupies a very special position at the very tip of the distribution of Hodge numbers. This manifold was found by one of the present authors (VB) by applying the methods that realise the symmetry of the covering space of the (1,4)-manifold as an operation on the toric polyhedron. More precisely, it was known that one of the polytopes of the Kreuzer-Skarke list is the 24-cell, a regular self-dual polytope that has 24 vertices and 24 three-faces. This polytope has a large symmetry group. There are three subgroups, $G$, with the property that 
$G$ acts freely and transitively on the vertices. One checks that this action on the vertices leads to a free action on the manifold. The most general polynomial invariant under $G$ is thus of the form $P{-}\vph Q$, where $P$ is the sum, with unit coefficients, of the monomials corresponding to the vertices of the polyhedron and 
$Q$ is the product of all the coordinates, and is the monomial corresponding to the interior point of the polytope. Since there is just one complex structure parameter, $h^{21}{\,=\,}1$, for the quotient, and since $\chi{\,=\,}0$ for the quotient, we also have $h^{11}{\,=\,}0$.
Full details of this construction may be found in \cite{Braun:2011hd}.

The reason for presenting the special geometries of these two manifolds together is that they are rather similar, even though there is no, currently known, direct relation between the corresponding \cy geometries. In each case there is a point of maximally unipotent monodromy, corresponding to the large complex
structure limit, there is no Gepner point, so no point with finite monodromy, and both of the special geometries have 6 \hbox{(hyper-)}conifold points. There are also singular points of the Picard-Fuchs equations for which
the monodromy is trivial. There are two such points for the (1,4)-manifold and one for the (1,1)-manifold. These are nonsingular points of the parameter space where the Yukawa couplings vanish.

Another matter that we find to be of interest has to do with the relation between two bases for the periods. For each of our two spaces we have a fourth order differential operator $\cL$. We denote by 
$\vph$, in each case, a coordinate for the special geometry, and choose $\vph$ such that the large complex structure limit corresponds to $\vph{\,=\,}0$. The indices for $\cL$, at this point, all vanish, so a basis of solutions for a neighbourhood  of $\vph{\,=\,}0$ has the form
\beq\label{FrobBasis}\begin{split}
\vp_0~&=~h_0 \\[3pt]
\vp_1~&=~\frac{1}{2\p\ii} h_0 \log\vph + h_1\\[3pt]
\vp_2~&=~\frac{1}{(2\p\ii)^2} h_0 \log^2\vph +\frac{2}{2\p\ii}h_1 \log\vph + h_2 \\[3pt]
\vp_3~&=~\frac{1}{(2\p\ii)^3} h_0 \log^3\vph + \frac{3}{(2\p\ii)^2} h_1 \log^2\vph +
\frac{3}{2\p\ii} h_2 \log\vph + h_3~,\\ 
\end{split}\eeq
where the $h_j$ are power series in $\vph$. As it stands, the basis is not unique\footnote{Since we may add multiples of $\vp_0$ to $\vp_1$, multiples of $\vp_0$ and $\vp_1$ to $\vp_2$, and, finally, multiples of $\vp_0$, $\vp_1$ and 
$\vp_2$ to $\vp_3$.}, but it becomes so if we require the boundary conditions $h_0(0){\,=\,}1$ and $h_j(0){\,=\,}0$. With these conditions, we refer to this basis as the Frobenius basis. We also assemble these periods into a vector
\beq
\vp~=~\begin{pmatrix}\vp_0\\ \vp_1\\ \vp_2\\ \vp_3\end{pmatrix}~.
\label{pivec}\eeq
We may also form another vector of periods $\P$ corresponding to a symplectic basis of $H^3(M,\IZ)$.
Interest then attaches to the constant matrix $\rho$ that relates these vectors.
\beqnn
\P~=~\rho\, \vp~.
\eeqnn
Now we can explain an observation that could have been made already with respect to the parameter space of the mirror quintic, but it arises here rather more dramatically because of the need for numerical computation. Consider the effect of continuing the vector $\vp$ around a loop that starts near 
$\vph{\,=\,}0$ and encloses one of the conifold points. With this loop we associate a monodromy matrix $m$
\beqnn
\vp\longrightarrow m\,\vp
\eeqnn
Since we do not know the periods in closed form, for the manifolds we consider here, we must use numerical methods to perform the analytic continuation. We can, for example, integrate the equations $\cL\vp{\,=\,}0$ numerically around a contour, that encloses the conifold point. We obtain in this way a numerical matrix $m$. For the (1,1)-manifold the matrix we obtain in this way is immediately seen to be a matrix of rational numbers of low height. For the (4,1)-manifold, by contrast, this is not so. Some of the elements of $m$ are clearly rational numbers, but others are not. Since we know that the special geometry prepotential contains a term, with the universal form, $\chi \z(3)/(2\p\ii)^3$, where $\chi$ is the Euler number of $M$, we hypothesize that the irrational elements of $m$ are (linear combinations of) powers of this number multiplied by rational numbers of low height. In other words the elements of $m$ take values in the field
\beqnn
\IK~=~\IQ\left[\frac{\chi \z(3)}{(2\p\ii)^3}\right]~,
\eeqnn
of rational numbers extended by $\chi \z(3)/(2\p\ii)^3$. The results of numerical computation are that this hypothesis is correct. The reason for including $\chi$ in this statement is that if $\chi{\,=\,}0$,
as it is for the (1,1)-manifold, then the assertion is that the elements of $m$ take values in $\IQ$. 

Now if we transport the period vector $\P$ around the same loop, we have
\beqnn
\P\longrightarrow M\,\P~,
 \eeqnn
with $M$ a matrix of integers. Since we also have
\beqnn
m~=~\rho^{-1} M \rho
 \eeqnn
it must be the case that $\rho$ takes values in $\IK$.

In the following two sections we study the special geometry of the (4,1) and (1,1) manifolds. We find the respective Picard Fuchs equations by first finding the series expansion of the fundamental period and then seeking the fourth order differential equation that it satisfies. Once we have the Picard Fuchs operator $\cL$ we are able to calculate the various monodromy matrices as indicated above. The relation between the Frobenius basis and integral basis is determined by choosing the matrix $\rho$ so that the monodromy matrices $\rho m \rho^{-1}$ are integral.

The calculation starts with the calculation of a sufficient number of coefficients $a_n$ that appear in the series expansion of the fundamental period.
\beqnn
\vp_0(\vph)~=~\sum_{n=0}^\infty a_n\, \vph^n~.
\eeqnn
Since the mirror manifolds form a 1-parameter family we expect $\vp_0$ to satisfy a fourth-order linear differential equation with regular singularities. We therefore seek a differential operator, with polynomial coefficients,
\beq
\cL~=~R_4(\vph)\,\vth^4 + R_3(\vph)\,\vth^3 + R_2(\vph)\,\vth^2 + R_1(\vph)\,\vth + R_0(\vph)~,~~
\text{where}~~~\vth~=~\vph\frac{d}{d\vph}~,
\label{eq:PFeqn}\eeq
that anihilates $\vp_0$. To this end we choose a maximal degree, $k_\text{max}$, for the polynomials 
$$
R_j(\vph)~=~\sum_{k=0}^{k_\text{max}} r_{jk}\,\vph^k
$$ 
and solve the linear equations, that result from requiring that $\cL\vp_0$ should vanish, for the coefficients 
$r_{jk}$. If $k_\text{max}$ is chosen too small then the only solution is that the $R_j$ vanish identically. We increase $k_\text{max}$ until we find a nontrivial solution. For the (4,1)-manifold we find 
$k_\text{max}=8$ and for the $(1,1)$-manifold we find $k_\text{max}=7$. In order to find these operators, and include some extra terms, as a check, we require the first 40 terms, say, in the expansion of 
$\vp_0$. A brute force calculation of this many terms of the series is difficult, so a more efficient method is required. For the cases we study we are able to calculate a sufficient number of coefficients by ad hoc methods. Once we have the Picard-Fuchs equation, however, we can compute many more coefficients $a_n$ by using the differential equation to find a recurrence relation.

Let us employ an multi-index notation and denote monomials that appear in the toric (Laurent) polynomial that defines the manifold by $t^v{\,=\,}\prod_j t_j^{v_j}$, in terms of four variables $t_j$. The exponents $v$ define a convex lattice polyhedron $\D$, with the origin $v{\,=\,}0$ as an interior point. With these conventions, one parameter families of \cy hypersurfaces in toric varieties, such as the two families we consider here, are described by the vanishing of polynomials of the form
\beq
1 - \vph \hskip-5pt\sum_{v\in\D\setminus\{0\}}\hskip-5pt t^v~.
\label{eq:DefiningPoly}\eeq

The exponents that appear in the defining polynomial define also a semigroup
\beq
\ccS~=~\left\{\sum_{v\in\D} n_v\,v ~\Big|~n_v\in\IN\right\}
\notag\eeq
\vskip-10pt
for which, by definition, all the coefficients $n_v$ are nonnegative.  For the Calabi-Yau hypersurfaces
that we will consider in this paper, the universal cover is a hypersurface in a particularly symmetric toric variety. In fact,  $\Delta$ will be the fundamental region of a lattice, that is, the translates of $\Delta$ tile $\ccM$. In that case we can also describe the vertices as the nearest neighbour lattice points of the origin and the semigroup $\ccS$ coincides with the lattice~$\ccM$.

Consider now the process of counting the number of lattice walks, for $\ccM$, from the origin to a lattice site $u$, in $n$ steps. The answer is given by the coefficient of $t^u \varphi^n$ in the multivariate generating series
\beqnn
\sum_{n=0}^\infty \Big(\varphi\sum_{v} t^{v}\Big)^n =~ \frac{1}{1 - \varphi \sum_{v} t^{v}}
\eeqnn
where the generating function is written in terms of four formal variables $t_j$, and the sum over $v$ is taken over 
$v\in\D\setminus\{0\}$. For a toric hypersurface defined by a polynomial of the form \eqref{eq:DefiningPoly} it can be shown that coefficients, $a_n$, in the series expansion of the fundamental period 
\beq
\vp_0(\vph)~=~\sum_{n=0}^\infty a_n\,\vph^n
\notag\eeq
are given by the coefficient of $\vph^n t^0$ in the expansion of 
\beq
\frac{1}{1 - \varphi \sum_{v} t^{v}}~.
\notag\eeq
Thus $a_n$ counts the number of lattice walks that return to the origin after $n$ steps and $\vp_0$ is the corresponding generating function.

We study the parameter space of the (1,4)-manifold in \SS2 and the (1,1)-manifold in \SS3. In \SS4 we return to the interpretation of $\vp_0$ as the generating function of lattice walks that return to the origin after $n$ steps.

The coefficients of the periods encode also interesting number-theoretic information about the underlying \cys. A knowledge of the coefficients provides, for example, a way to calculate the zeta-functions for these manifolds. This is an interesting topic to which we will return elsewhere. 

The periods of a \cym play an essential role in the description of the manifold and of its parameter space. The periods, in turn, satisfy Picard-Fuchs equations. The question arises as to which differential operators are associated with the parameter spaces of \cys. Almqvist {\sl et al.\/}~\cite{Almkvist:2005} have formulated conditions that set out when a differential operator is of `\cy type' and have assembled a table of operators that satisfy the conditions. If these conditions are satisfied, then the yukawa coupling, computed from the periods, has a $q$-expansion
\beq
y~=~y_0 + \sum_{k=1}^\infty \frac{n_k k^3 q^k}{1-q^k}
\notag\eeq
with integer coefficients $y_0$ and $n_k$. In some cases the differential equation and the periods are known to correspond to a \cym. In other cases this is not known. It was observed by Guttman~\cite{Guttmann:2009} and by Broadhurst~\cite{Guttmann:2010nj} that certain generating functions for lattice walks satisfy differential equations of \cy type and two of these functions and equations correspond to the manifolds that we discuss here. These differential equations are listed in the table of \cite{Almkvist:2005}. Thus the differential equations were already known and the expansions for the yukawa couplings were known up to an overall multiplicative integer. Here we derive the periods and the differential equation from the manifolds in question and study the monodromies of the periods around the singularities of the moduli space. The observations that we make here about the occurence of the quantity $\z(3)/(2\p\ii)^3$ in the matrix that relates the Frobenius basis of the periods to the integral basis, and in the monodromy matrices for the Frobenius basis, could have been made already in relation to the quintic \cym. In the context of the monodromy matrices for the operators of \cy type, the ubiquity of $\z(3)/(2\p\ii)^3$ was noted in the extensive calculations of Hoffman~\cite{hofmann2013monodromy}.
\newpage
\section{The Special Geometry of the (4,1)-Manifold}\label{sec:41manifold}
\vskip-10pt
\subsection{Fundamental Period}\label{sec:41period}
From~\cite{Braun:2009qy} we have that the fundamental period is of the form discussed in the Introduction. It is given by the following integral representation
\beqnn
\vp_0(\vph)~=~\frac{1}{(2\p\ii)^4}\int\frac{\dd^4 t}{t_1 t_2 t_3 t_4\, \big(1 - \vph f(t)\big)}
\eeqnn
where 
\begin{equation}
  \label{eq:Tdef}
  f(t)~=~T(t_1,t_2) + T(t_3,t_4) ~~~\text{with}~~~T(t_1,t_2)~=~t_1 + \frac{1}{t_1} + t_2 +
  \frac{1}{t_2} + \frac{t_1}{t_2} + \frac{t_2}{t_1}~,
\end{equation}
and the contour, for the integral, consists of a product of four loops enclosing the poles $t_j{\,=\,}0$.
From the integral we see that
\beqnn
\vp_0(\vph)~=~\sum_{n=0}^\infty \vph^n\,\big[f(t)^n\big]_{t^0}~,
\eeqnn
with $\big[f(t)^n\big]_{t^0}$ denoting the constant term in the expansion of $f(t)^n$. In order to compute the coefficients $[f(t)^n]_{t^0}$ efficiently we set 
$T_k = \big[T(t_1,t_2)^k\big]_{t^0}$ and note that
$$
\big[f(t)^n\big]_{t^0}~=~\sum_{r=0}^n \binom{n}{r}\, T_{n-r}T_r ~=~\begin{cases}
\displaystyle 2\sum_{r=0}^{\frac{1}{2}(n-1)}\!\binom{n}{r}\, T_{n-r}T_r~;& \text{if $n$ is odd}\\[18pt] 
\displaystyle 2~\sum_{r=0}^{\frac{n}{2}-1}\;\binom{n}{r}\, T_{n-r}T_r + 
\binom{n}{\frac{n}{2}}\,T_{\frac{n}{2}}^2~; & \text{if $n$ is even}~.\end{cases}
$$
We can calculate the quantities $T_k$ explicitly by collecting powers of $t_1$
$$
T(t_1,t_2)~=~t_1\left( 1 + \frac{1}{t_2}\right) + \left( t_2 + \frac{1}{t_2}\right) + \frac{1}{t_1}(1+t_2)
$$
and picking out the terms in the expansion of $T(t_1,t_2)^k$ that are independent of $t_1$ and then of $t_2$.
In this way we find
\beq
T_k~=~\sum_{\raisebox{-3pt}{$\scriptstyle r=0$}}
^{\raisebox{3pt}{$\scriptstyle \lfloor \frac{k}{2}\rfloor$}}
\hskip5pt\sum_{s=\max\left(0,\lceil \frac{k-3r}{2}\rceil\right)}^{\min\left(k-2r,\lfloor\frac{k-r}{2}\rfloor\right)}
\frac{k!\, (2r)!}{(r!)^2 \, s!\, (3r + 2s - k)!\, (k - r -2s)!\, (k - 2r -s)!}~.
\label{Tcoeffs}\eeq
It is easy, proceeding in this way, to compute several hundred terms, say, of the series. 
The first few terms in the expansion of $\vp_0$ are given by
\begin{equation}
  \label{pidef}
  \begin{split}
    \vp_0(\vph)~&=~ 1 + 12\, \vph^2 + 24\, \vph^3 + 396\, \vph^4 + 2160\, \vph^5 + 23160\, \vph^6 +
    186480\, \vph^7\\[3pt]
    &\hskip27pt + 1845900\, \vph^8 + 17213280\, \vph^9 
    + 171575712\, \vph^{10} + O(\vph^{11})
  \end{split}
\end{equation}
In terms of lattices, the $12$ vertices of the dual polytope arrange
into the nearest neighbours of two copies of the two-dimensional
hexagonal (a.k.a.\ triangular) lattice. This much is clear from the
description of the $(4,1)$-manifold as a hypersurface in the product
of two toric $\dP_6$ surfaces. The $T_k$ count closed length-$k$ walks
in the hexagonal lattice. There is no closed form (not involving a
double sum) for the coefficients, but we cannot resist remarking that there are interesting expressions that are known ordinary and exponential generating functions~\cite{oeisA002898}:
\begin{equation*}
  \begin{split}
    \sum_k T_k x^k ~&=~ 
    \sum_k \frac{(3k)!}{k!^3} x^{2k} (1+2x)^k ~=~
    {}_2F_1\big(\tfrac{1}{3},\, \tfrac{2}{3};\, 1;\, 27 x^2 (1+2x)\big),
    \\[5pt]
    \sum_k \frac{T_k}{k!} x^k ~&=~\; 
    I_0(2x)^3 + 2 \sum_{k=1}^\infty I_k(2x)^3.
  \end{split}
\end{equation*}
Where ${}_2F_1$ denotes the hypergeometric function and the $I_k$ are Bessel functions of imaginary argument.
The exponential generating function is multiplicative over lattice
products, hence we can write the exponential generating function of
the fundamental period using Bessel functions as
\begin{equation*}
 E_{\vp_0}(\vph) ~=~\sum_{n=0}^\infty \frac{a_n}{n!}\, \vph^n ~=~
  \left( I_0(2\vph)^3 + 2 \sum_{k=1}^\infty I_k(2\vph)^3 \right)^2.
\end{equation*}
Another way of thinking about the closed walks in the hexagonal
lattice is as closed paths where the steps are sixth roots of
unity. In fact, a walk whose steps are \emph{twelfth} root of unity
closes if and only if the even and odd powers of the fundamental root
close separately. So, if $\b$ is a primitive twelfth root of unity, we have
\begin{equation*}
  \sum_{j=0}^{11} k_j \b^j = 0
  \quad \Leftrightarrow\quad
  \sum_{j=0}^{5} k_{2j} \b^{2j} = 0
  \quad\text{and}\quad
  \sum_{j=0}^{5} k_{2j+1} \b^{2j+1} = 0.
\end{equation*}
Therefore, we can also identify the fundamental period as the
(ordinary) generating functions of closed walks using $12$-th roots of
unity~\cite{Labelle:2011}, leading to the more symmetric formula
\begin{equation*}
  \varpi_0(\vph) ~=
  \sum_{
    \begin{subarray}{c}
      k_0 + \cdots + k_{11} = n\\[2pt]
      k_0 k_{10} \,=\, k_4 k_6 \\[2pt]
      k_1 k_{11} \,=\, k_5 k_7 \\[2pt]
      k_2 k_4 \,=\, k_8 k_{10} \\[2pt]
      k_3 k_5 \,=\, k_9 k_{11}
    \end{subarray}
  }
  \frac{n!}{k_0! k_1! \cdots k_{11}!} \vph^n
\end{equation*}
\newpage

\subsection{Picard-Fuchs Differential Equation}
We seek a fourth order differential equation that the fundamental period satisfies, as indicated in the Introduction, and increase $k_\text{max}$ until we come to $k_\text{max}=8$. This process yields the following coefficient functions $R_j$:
\beq\begin{split}
R_4&= 69120\,\left(\vph -\smallfrac32\right)^2 \left(\vph +\smallfrac14\right)
\left(\vph +\smallfrac15\right) \left(\vph +\smallfrac16\right) 
\left(\vph -\smallfrac{1}{12}\right) \left(\vph -\smallfrac14\right) \left(\vph -\smallfrac13\right) \\[15pt]  R_3&= 8\vph \smash{\left(\vph - \smallfrac32\right)} \left(86400 \vph^6-158976 \vph^5-1512 \vph^4+15964 \vph^3+160 \vph^2-345 \vph -5\right)\\[15pt]
R_2&= \vph  \big(2419200 \vph^7-8581248 \vph^6+7771104 \vph^5+274360 \vph^4-552220
   \vph^3-5250 \vph^2\\
&\hskip11.5cm+6917 \vph +39\big) \\[15pt] 
R_1&= \vph  \big(3456000 \vph^7-12745728 \vph^6+12372480 \vph^5+166288 \vph^4-679952
   \vph^3-1584 \vph^2\\
&\hskip11.5cm+5532 \vph +9\big) \\[15pt]
R_0&= 48 \vph^2 \left(34560 \vph^6-130464 \vph^5+132120 \vph^4+284 \vph^3-6182\vph^2
+9 \vph +36\right) \\
\end{split}\notag\eeq
Notice that, apart from the factor of $\left(\vph-\smallfrac32\right)^2$, we recognise the factors of $R_4$ from Table~7 of~\cite{Braun:2009qy}. From the ratio $R_3/R_4$ we may compute the Yukawa coupling
\begin{equation*}
\begin{split}
y_{\vph\vph\vph}~&=~\frac{1}{\vph^3}\,\exp\left( -\frac12 \int\! \frac{d\vph}{\vph}\, \frac{R_3(\vph)}{R_4(\vph)} \right)\\[10pt]
&=~\text{const.}\times\frac{\vph -\smallfrac32}{\vph^3 \left(\vph +\smallfrac14\right)
\left(\vph +\smallfrac15\right) \left(\vph +\smallfrac16\right) 
\left(\vph -\smallfrac{1}{12}\right) \left(\vph -\smallfrac14\right) \left(\vph -\smallfrac13\right)}~.
\end{split}
\end{equation*}
Notice that the coupling has a zero at $\vph=\smallfrac{3}{2}$.

Since we now have the differential equation we may compute the indices at the singular points. These are summarised by the Riemann symbol, which records the orders of the vanishing of the four solutions of the Picard--Fuchs equation at the singularities of the equation.
$$
\cP\left\{\begin{array}{rccrrrcccc}
0 &~\infty~ &~\frac32~ &-\frac14 &-\frac15 &-\frac16 &~~\frac{1}{12}~ &~\frac14~ &~\frac13~&\\[3pt]
0 & 1           & 0             & 0         & 0          & 0          & ~0                      & 0             & 0            &\\
0 & 2           & 1             & 1         & 1          & 1          & ~1                      & 1             & 1            & \vph\\
0 & 3           & 3             & 1         & 1          & 1          & ~1                      & 1             & 1            &\\
0 & 4           & 4             & 2         & 2          & 2          & ~2                      & 2             & 2            &\\[3pt]
\end{array}\right\}
$$
Note first that all four indices at $\vph{\,=\,}0$ vanish so this is a point of maximal unipotent monodromy. This confirms our identification of this point as the large complex structure limit. The method of Frobenius yields a basis of solutions for the operator $\cL$ as in \eqref{FrobBasis}.
The procedure is readily adapted to computer calculation. First, given $\cL$, we solve the equation 
$\cL\vp_0=0$ for the coefficients of the series $h_0$, then we solve the equation $\cL\vp_1=0$ for the coefficients of the series $h_1$, and so on. 

We form the periods $\vp_j$ into a vector, as in \eqref{pivec}, and consider the monodromy 
$\vp \to m_0\,\vp$ going anti-clockwise in a small circle about the origin. This monodromy is generated by 
$\log\vph \to \log\vph + 2\p\ii$, and in this way we see that
$$
\label{eq:MUM}
m_0~=~\begin{pmatrix}
1 & 0 & 0 & 0\\
1 & 1 & 0 & 0\\
1 & 2 & 1 & 0\\
1 & 3 & 3 & 1 \end{pmatrix}
$$
Consider next the monodromy about one of the six points that correspond to the indices $\{0,1,1,2\}$ which, given the indices and the type of singularity recorded in Table~7 of~\cite{Braun:2009qy}, we expect to be of conifold type. That is we expect $m$ to have the form
$$
m~=~\one + a~,
$$
with $a^2{\,=\,}0$ so that $m^n = \one + n a$. We can say more, since we expect $a$ to be a matrix of rank 1 and therefore to have the structure
$$
a_{ij}~=~u_i v_j ~~~\text{with}~~~ u_j v_j~=~0~,
$$
the latter condition following from the fact that $a^2=0$. Mathematica provides a function that, given suitable initial conditions, integrates a differential equation along a line segment and returns an interpolating function. Four such line segments may be concatenated to form a rectangle. So, starting from a point near $\vph=0$, where the values of $\vp$ and its derivatives may be computed from \eqref{pidef}, we may integrate the  Picard-Fuchs equation around a rectangle that encloses some of the singular points in order to find the monodromy matrices. 

Consider the monodromy matrix that corresponds to the singularity at $\vph=1/12$. The numerical procedure just described returns the matrix\footnote{It is straightforward to perform the computations to 70 figures, say. But we show a simplified output.}
$$
m_{\frac{1}{12}} -\one ~=~
\left(\begin{array}{cccc}
    \,0.348914\, \ii & -6~~                & 0 & -36~~ \\
                        0 & 0                      & 0 & 0 \\
~~ 0.0193841\,\ii & -0.333333~~~ & 0 & -2~~ \\
    -0.00338169     & -0.0581523\,\ii & 0 & -0.348914\,\ii
\end{array}\right)\raisebox{-25pt}{.}
$$
Let us denote the $(1,1)$-component of the matrix above by $\ii\a$. We see that the matrix
has indeed the form $u_i v_j$ since the second, third and fourth rows are obtained from the first by multiplying by $0,\, 1/18$ and $\ii\a/36$, respectively.  
Let us define now two antisymmetric matrices which prove useful in the following
\beq
\S~=~\left(\begin{array}{rrrr}
       0 &      \+ 0 &     \+ 1 & \+ 0\\
       0 &          0 &          0 &     1\\
\!\! -1 &          0 &          0 &      0\\
        0&        -1 &          0 &      0
\end{array}\right)
~~~\text{and}~~~
\s~=~\frac{1}{Y_{111}} 
\left(\begin{array}{rrrr}
       0 &      \+ 0 &      \+0 & - 6\\
       0 &          0 &          2 &     0\\
       0 &         -2&          0 &     0\\
        6&         0 &          0 &     0
\end{array}\right)
\raisebox{-25pt}{,}
\label{eq:symplectic}\eeq
where we write an explicit factor of the classical yukawa coupling (=18 for this manifold) in our definition of $\s$ to show the dependence on the yukawa coupling for different manifolds.
We find, in fact, that $m_\frac{1}{12}$ has the form
$$
m_{\frac{1}{12}}~=~\one - 12\, u\, u^T\!\s^{-1} ~~~\text{with}~~~ 
u~=~\begin{pmatrix} 1 \\[2pt] 0\\[2pt] 1/18 \\[2pt] \ii\a/36 \end{pmatrix}\raisebox{-25pt}{.}
$$
The prepotential contains a term $\ii\z(3)/(2\p)^3$ and this suggests that we try to identify the number $\a$ in terms of this. We find
$$
\a~=~72\,\frac{\z(3)}{(2\p)^3}~.
$$
The monodromies about the other conifold points have an analogous structure. In each case the matrix takes the form
\beq
m_{ij}~=~\d_{ij} - \mu\, u_i (u^T\!\s^{-1})_j~.
\label{matrixm}\eeq
We record the coefficients $\mu$ and vectors $u$ in \tref{MonodromyTable}.
\begin{table}[htb]
  \def\str{\vrule height23pt depth16pt width0pt}
  \begin{center}
    \newcolumntype{C}{>{~$}c<{$~}}
    \newcolumntype{R}{>{~$\displaystyle}r<{$~}}
    \newcolumntype{L}{>{~~$}l<{$~~}}
    \begin{tabular}{| R | C | C | C | L | L |}\hline
     \str \vph~ & \text{number} & \text{singularity} & ~~\mu~~ & \hfill u^T\hfill{} & \hfill U^T\hfill{} \\ 
\hline\hline
\str -\frac{1}{4} & (ii),\,(x) & 1\, g_4,\,1\,g_4^2& 12 
& \left(3, 1, \frac{7}{18}, \frac{1}{6} + \frac{\ii\a}{12} \right) 
&(1, -2,\, 3,\, 1) \\ 
\hline
\str -\frac{1}{5} & (vi) & 1 & 1 
& \left(12, 5, \frac{7}{3}, \frac{7}{6} + \frac{\ii\a}{3}\right)
& (6, -15,\, 12,\, 5) \\
\hline
\str -\frac{1}{6} & (iv) & 1\, g_3 &  3 
& \left(4, 2, \frac{10}{9}, \frac{2}{3} + \frac{\ii\a}{9}\right)
& (3, -8,\, 4,\, 2) \\
\hline
\str  \frac{1}{12} & (iii) & 1\,\goth & 12 & \left(1,\, 0,\, \frac{1}{18},\, \frac{\ii\a}{36}\right)
& (0,\, 0,\, 1,\, 0) \\
\hline
\str  \frac{1}{4} & (ix) & 1\, g_4^2 & 2 
& \left(6, 1, \frac{1}{3}, \frac{1}{6} + \frac{\ii\a}{6}\right)
& (1,\, 0,\, 6,\, 1) \\
\hline
\str  \frac{1}{3} & (v) & 1\, g_3 &  3 
& \left(4, 1, \frac{1}{3}, \frac{1}{6} + \frac {\ii\a}{9}\right)
& (1, -1,\, 4,\, 1)\\ 
\hline
\end{tabular} 
\capt{5.6in}{MonodromyTable}{The vectors $u$ corresponding to the
  monodromy matrices for each of the hyperconifold points. The vectors
  $U$ are the same vector in the integral symplectic homology basis,
  see \sref{sec:HZZ}. The ``number'' column refers to the labelling
  of the components of the discriminant locus given
  in~\cite{Braun:2009qy} and the ``singularity'' column records the
  number and type of the hyperconifold
  singularities~\cite{Davies:2011is}. Thus $1\,\goth$, for example,
  refers to a single hyperconifold point that is the $\goth$-quotient
  of a node, while $1\, g_4,\,1\,g_4^2$, refers to one hyperconifold
  that is a $g_4$-quotient and one that is a $g_4^2$-quotient.}
\end{center}
\end{table}
The points $\vph=3/2$ and $\vph=\infty$ are singularities of the Picard-Fuchs equation though the manifold is smooth for these values. Computation reveals that the monodromy around these points is trivial
$$
m_\frac32~=~m_\infty~=~\one~.
$$
The fact that $m_\frac32$ is trivial is checked by direct computation. The triviality of the monodromy about 
$\infty$ is then a consequence of the fact that the total monodromy around all other singularities is trivial and this amounts to the identity
$$
m_{-\frac14}\, m_{-\frac15}\, m_{-\frac16}\, m_0\; m_\frac{1}{12}\, m_\frac14\, m_\frac13 ~=~\one~. 
$$
Owing to the monodromy the periods $\vp_j$ are multiple valued functions. We may render them single valued by cutting the $\vph$-plane from $\vph=-1/3$ to $\vph=1/4$ and taking the expressions \eqref{pidef} to hold in the region $|\vph|<1/12,~\im(\vph)\geq 0$.
\begin{figure}[H]
\centering
\framebox[5.2in][c]{\vrule width0pt height 2.15in depth 0in\includegraphics[width=5in]{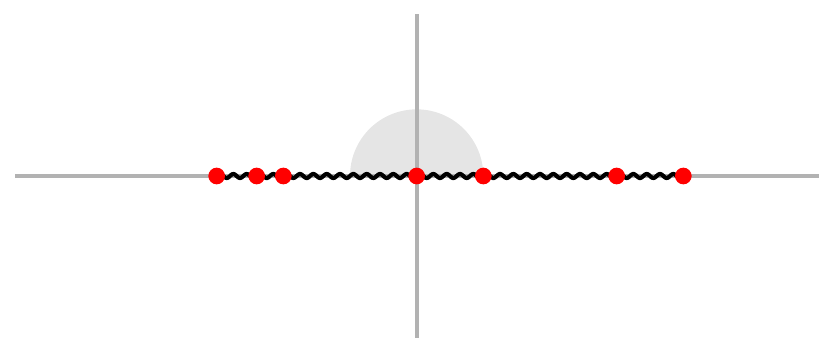}}
\capt{5in}{figPhiPlane}{
The $\vph$-plane cut from $-1/4$ to $1/3$ and showing the singularities of 
$\vp(\vph)$ at $\vph=-\frac14,-\frac15,-\frac16, 0, \frac{1}{12}, \frac14, \frac13$. The shaded area
corresponds the region where the expression \eqref{pidef} holds.}
\end{figure}
\subsection{Integral Homology Basis and the Prepotential}
\label{sec:HZZ}
We turn now to the task of finding an integral basis for the homology
and the prepotential, $\Fprepotential$, for the mirror of $\YG$. For the
perturbative part of the prepotential we take
$$
\Fpert~=~-\frac{1}{3!}\, Y_{abc} \frac{z^a z^b z^c}{z^0} ~~~\text{with}~~~
Y_{111}~=~18~,
$$
which is written in terms of the special coordinates $(z^0,z^1)$ and the value given for $Y_{111}$ follows from an intersection calculation.

In general, when there are $n$ parameters, the indices run over the range $a,b,c = 0,\ldots, n$. For $i,j,k=1,\ldots,n$ the $Y_{ijk}$ are the topological intersection numbers
$$
Y_{ijk}~=~\int_\cM \! e_i\, e_j\, e_k~,
$$
with the $e_i$ a basis for $H^2(\cM,\IZ)$. The remaining coefficients $Y_{0jk}$, $Y_{00k}$ and 
$Y_{000}$ depend on a choice of symplectic basis for $H^3(\cW,\IZ)$. In the following we will make reference to the Chern classes of $\cM$ which we denote by $c_r$. The term $Y_{000}$ is believed to be related to the four loop term for the $\beta$-function for $(2,2)$ supersymmetric sigma-models on \cy threefolds and to be, in all cases, such that by means of an integral symplectic transformation it may be brought to the form
\beq
Y_{000}~=~-3\frac{\z(3)}{(2\p\ii)^3}\,\chi(X)~.
\label{ChernThree}\eeq
Where $\chi=\int c_3$ denotes the Euler number. It has been observed also that the terms $Y_{00k}$ can also be brought, by means of an $\Sp (n+1,\IZ)$ transformation, to the form
\beq
Y_{00k}~=~-\frac{1}{12}\int_\cM\! c_2\, e_k~.
\label{ChernTwo}\eeq
Given the last two relations it would seem intuitive that the terms $Y_{0jk}$ can be brought to a form
proportional to
\beq
\int_\cM \! c_1\,e_j\,e_k
\label{ChernOne}\eeq
and hence be set to zero. This cannot, however, be true in all cases since the prepotential for the mirror quintic furnishes a counterexample. In that case the quantity $Y_{011}$ takes values that are half an odd integer and the values can only be changed by an integer, by an $\Sp(4,\IZ)$ transformation. For the manifold that we are considering here we will, nevertheless, see that it is possible to take $Y_{011}=0$.

From the full prepotential, $\Fprepotential$, we form the period vector
$$
\P~=~ \begin{pmatrix} 
\displaystyle \pd{\Fprepotential}{z^0} \\[12pt] 
\displaystyle \pd{\Fprepotential}{z^1} \\[10pt] 
z^0 \\ 
z^1 \end{pmatrix}\raisebox{-40pt}{.}
$$
Both $\P$ and $\vp$ are vectors of periods so are related by a matrix of constants $\rho$
$$
\P~=~\rho\, \vp~.
$$
We identify $(z^0,\, z^1) = (\vp_0,\,\vp_1)$ and set
\beq
t~=~\frac{z^1}{z^0}~=~\frac{1}{2\p\ii}\log\vph + \frac{h_1(\vph)}{h_0(\vph)}~.
\label{tdef}\eeq
The series $h_1$ has been defined such that $h_1(0)=0$, so as $\vph\to 0$ we have 
$t = \frac{1}{2\p\ii}\log\vph + \cO(\vph)$. The asymptotic form of the period vectors, as $\vph\to 0$ is enough to relate 
$\rho$ to the constants that appear $\Fpert$. We have
$$
\P~\sim~ 
\begin{pmatrix} 
\+\frac{1}{6} t^3\,Y_{111} - \frac{1}{2} t\, Y_{001}-\frac{1}{3}\, Y_{000}\\[8pt]
-\frac{1}{2} t^2\, Y_{111} - \hphantom{\frac{1}{2}}t\, Y_{011} - \frac{1}{2}\, Y_{001}\\[5pt]
1\\[5pt]
t 
\end{pmatrix}
~~~\text{and}~~~
\vp~\sim~\begin{pmatrix}
1 \\[5pt] t \\[5pt] t^2 \\[5pt] t^3 \end{pmatrix}\raisebox{-25pt}{.}
$$
Thus
$$
\rho~=~ \left( \begin{array}{ccrr@{\hskip13pt}}
- \frac{1}{3} Y_{000} & -\frac{1}{2} Y_{001}       & 0   &\+{\hskip13pt}\frac{1}{6} Y_{111}{\hskip-13pt}\\[3pt]
 -\frac{1}{2} Y_{001} & - \hphantom{\frac{1}{2}}Y_{011} & -\frac{1}{2} Y_{111}{\hskip-13pt} & 0 \\[3pt]
 1 & 0 & 0 & 0 \\[3pt]
 0 & 1 & 0 & 0
\end{array} \right)\raisebox{-30pt}{.}
$$
Every matrix $\rho$, of this form, has the property that
$$
\rho\s\rho^T~=~\S~.
$$ 
Note also that any monodromy matrix of the form \eqref{matrixm} preserves the matrix $\s$
$$
m \s m^T~=~\s~.
$$
In the $\P$ - basis we have monodromy matrices
\beq
M~=~\rho m \rho^{-1}~=~\one + \mu\, U U^T\S ~~~\text{with}~~~U~=~\rho u
\label{IntegralMonodromy}\eeq
and these satisfy
$$
M\,\S\, M^T~=~\S~.
$$
Thus the matrices $M$ are symplectic. We need to check that they are also integral. The matrix 
$M_\frac{1}{12}$ has, among its components, the quantities $(\ii\a-4Y_{000})$ and 
$\frac43 (\ii\a-4Y_{000})^2$. For these to be integral we must have 
$$
Y_{000}~=~\frac{\ii\a}{4} + \frac32\, C~~~\text{with}~~~C\in\IZ~.
$$
The integrality of $M_0$ requires both $Y_{001}$ and $Y_{011}$ to be integral. Another constraint comes from $M_\frac13$, which contains the term $\frac32(Y_{001}-1+4C)$. Thus 
$$
Y_{001}~=~-1 + 2B ~~~\text{with}~~~ B\in\IZ~.
$$
For uniformity we write $Y_{011}=A$. With $A$, $B$ and $C$ integral all the matrices $M$ are integral. 

The matrix
$$
S~=~\begin{pmatrix}
 1 & 0 & C' & B'\\
 0 & 1 & B' & A' \\
 0 & 0 &  1  &  0 \\
 0 & 0 &  0  &  1  
\end{pmatrix}
$$
is integral and symplectic if $A'$, $B'$ and $C'$ are integral and 
$$
S\rho~=~\left( \begin{array}{ccrr}
 \frac{\ii\a}{12} - \frac12 C  + C' & \frac12 - A + A' & 0 &~~~~ 3\\[5pt]
\frac12 - A + A' &  - B + B' & -9 & 0 \\[5pt]
 1 & 0 & 0 & 0 \\[5pt]
 0 & 1 & 0 & 0
\end{array} \right)
$$
The upshot is that, by means of an $\Sp(4,\IZ)$ transformation that, moreover, preserves the coordinates 
$(z^0,\, z^1)$, we may set $A=B=0$ and also remove the integer part of $C/2$. Although it does not follow from the integrality of $M$ we will nevertheless assume that $C$ is even so that it may be removed by a symplectic transformation in conformity with our previous observations concerning $Y_{000}$.
Thus we take $\rho$ to be given by
$$
\rho~=~\left( \begin{array}{rrrr}
- \frac{\ii\a}{12} & \+ \frac12  & 0 &\+ 3\\[3pt]
\frac12  & 0 & -9 & 0 \\[3pt]
 1 & 0 & 0 & 0 \\[3pt]
 0 & 1 & 0 & 0
\end{array} \right)\raisebox{-28pt}{.}
$$

In summary we may take
$$
Y_{111}~=~18~,~~~Y_{011}~=~0~,~~~Y_{001}~=~-1~,~~~
Y_{000}~=~18\,\frac{\z(3)}{(2\p\ii)^3}~.
$$
For our manifold we have $\chi(\YG)=-6$ and $\int_{\YG} c_2\, e_1 = 12$, so the these values for the constants are in conformity with the forms \eqref{ChernThree}, \eqref{ChernTwo} and \eqref{ChernOne}.
With these values for the constants the perturbative part of the superpotential assumes the very simple form
$$
\Fpert~=~ - 3\frac{(z^1)^3}{z^0}+\frac12\, z^0 z^1 - 3\frac{\z(3)}{(2\p\ii)^3}\, (z^0)^2 ~.
$$

Let us return to the form of the monodromy matrices. In writing the matrices in the form 
\eqref{IntegralMonodromy} there is a freedom to transfer a numerical factor between the coefficient $\mu$ and the vector $U$. We fix the normalization of $U$, up to sign, by requiring that the components of $U$ be integral and have no common factor. The coefficients and vectors that are presented in \tref{MonodromyTable} are normalised in this way. Note that, with the possible exception of the monodromy about $\vph=-1/4$, the coefficients $\mu$ correspond to the `equivalent number of conifolds' with a $\goth$-hyperconifold counting as 12 conifold points and a $g_3$ hyperconifold counting as 3. The singularity for $\vph=-1/4$ corresponds to a $g_4$-hyperconifold and a $g_4^2$-hyperconifold so, from this point of view, ought to count as six conifold points instead of 12. The singularity at $\vph=-1/4$ is, however, special in that this singularity corresponds to the intersection of the $\vph$-plane with the two components $(ii)$ and $(x)$ of the discriminant locus. The component $(ii)$ corresponds to the hyperconifold singularities while the component $(x)$ is fixed by a possible $\IZ_2$ automorphism of the two parameter space $\G$.

\begin{figure}[p]
  \centering
  \includegraphics[width=\linewidth]{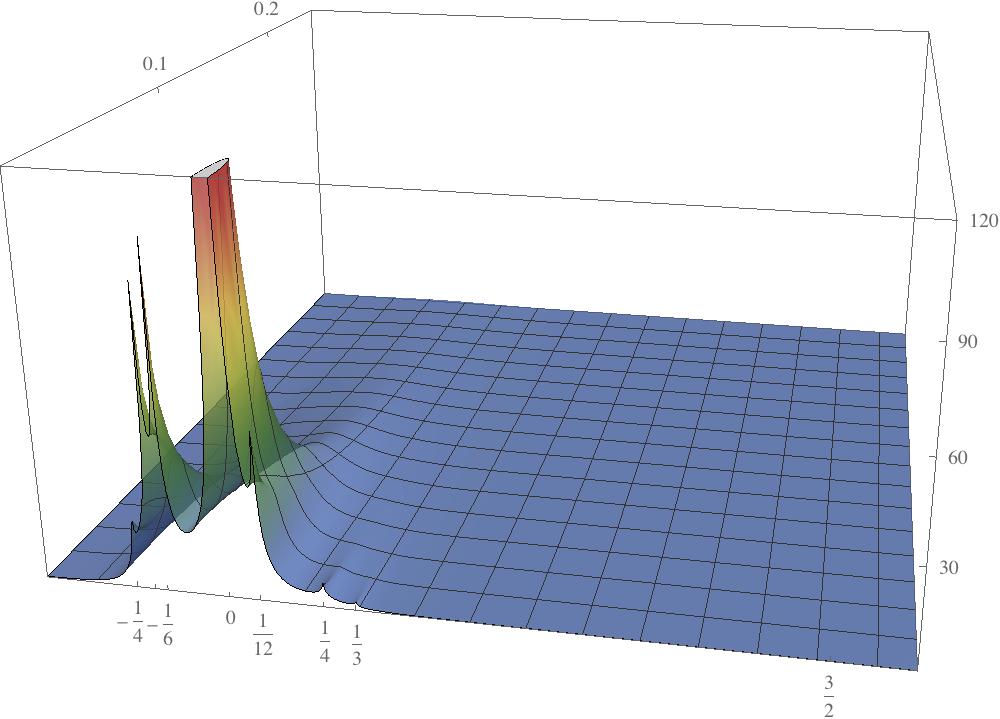}\\
  \vspace{0.1cm}
  \includegraphics[width=0.95\linewidth]{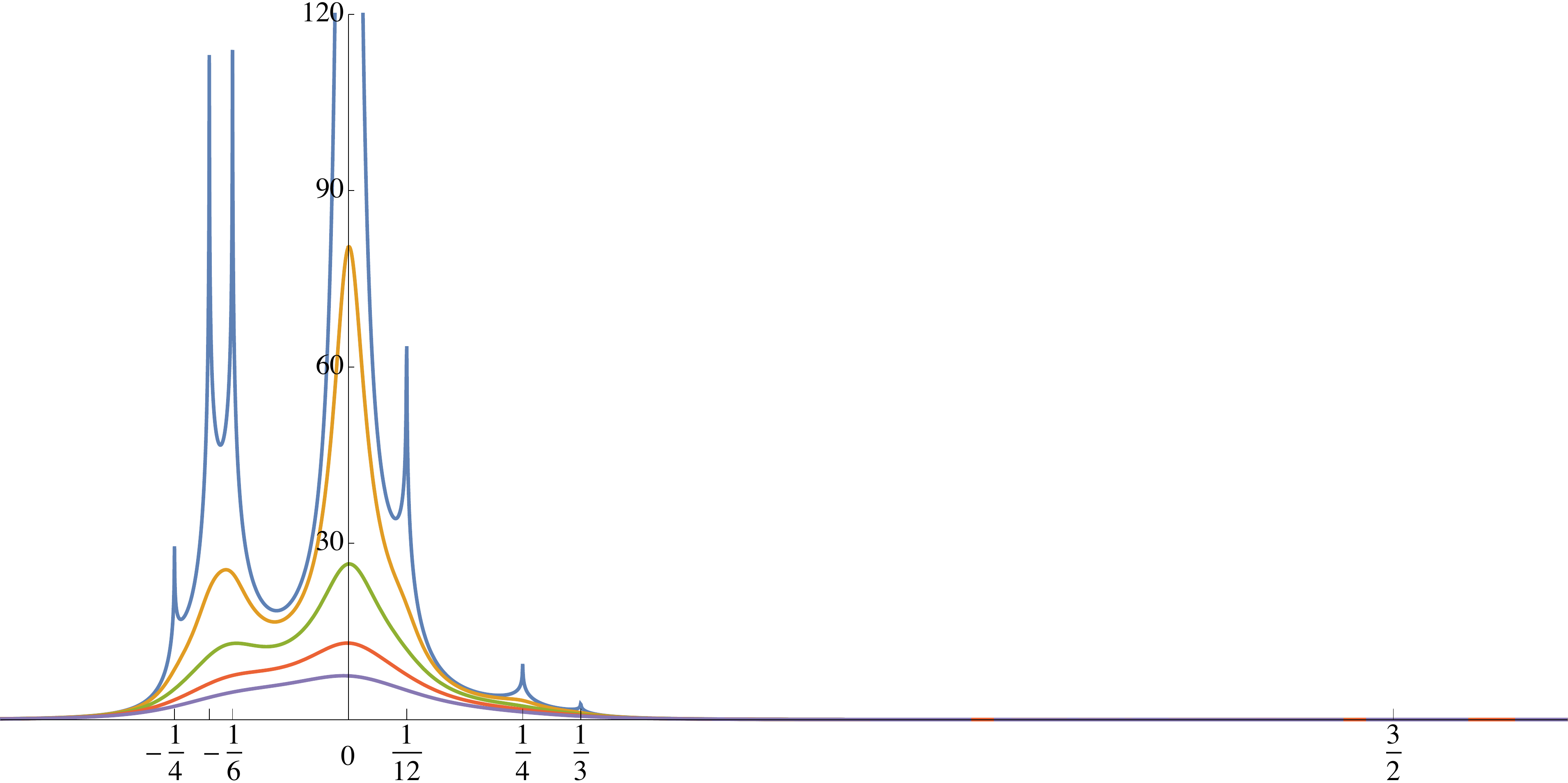}\\
  \vspace{10pt}
  \caption{The metric $g_{\vph\bar\vph}$ on the complex structure
    moduli space. Top: plot over a region of the complex plane.
    Bottom: restriction to a real interval (blue) and its shift to
    constant imaginary value (remaining colors).}
  \label{fig:metric}
\end{figure}
We have determined the matrix $\rho$, hence $\P$. The special geometry of the complex structures of the mirror of $\YG$ is now calculable in a straightforward way. The K\"ahler potential of the metric on the parameter space, for example, is given by
$$
\ee^{-K}~=~-\ii\,\P^\dag\S\,\P ~=~ -\ii\,\vp^\dag \rho^\dag\S\rho\,\vp
~=~ \frac{12\,\z(3)}{(2\p)^3} |\vp_0|^2 +\ii\,\vp^\dag\s^{-1}\vp~.
$$
Since the complex structure space is one-dimensional, there is only a
single nonvanishing metric component $g_{\vph\bar\vph}$. We plot its
behavior in \fref{fig:metric}.

\subsection{The Yukawa coupling}
The yukawa coupling, in the symplectic basis, is given by the expression
$$
y_{ttt}~=~\frac{1}{1440\,\vp_0(\vph)^2 }\,
\frac{\left(\vph -\frac{3}{2}\right)} {\vph^3 \left(\vph -\frac{1}{3}\right) \left(\vph-\frac{1}{4}\right) 
\left(\vph -\frac{1}{12}\right) \left(\vph +\frac{1}{4}\right)\left(\vph+\frac{1}{5}\right) 
\left(\vph +\frac{1}{6}\right)}
\left(\frac{1}{2\p\ii}\frac{d\vph}{dt}\right)^3~,
$$
an overall scale having been chosen such that $y_{ttt}\to 18$ as $\vph\to 0$. We invert the relation 
\eqref{tdef} to write $\vph$ as a series in the variable $q=\ee^{2\p\ii t}$. 
$$
\vph~=~q - q^2 - 14\, q^3 + 9\, q^4 + 146\, q^5 - 406\, q^6 - 2967\, q^7 + 2055\, q^8 + 9100\, q^9 + 
\cdots
$$
In this way we obtain a series expansion for $y_{ttt}$ which has the form of an instanton expansion
$$
y_{ttt}~=~18 + \sum_{k=0}^\infty \frac{n_k\, k^3\,q^k}{1 - q^k}~,
$$
with the first few instanton numbers given in the table. 
\begin{table}[H]
\vspace{20pt}
\begin{center}
\def\str{\vrule height14pt depth5pt width0pt}
\newcolumntype{C}{>{~$}c<{$~}}
\newcolumntype{R}{>{~$}r<{$~}}
\begin{tabular}{|C|R|}
\hline
\str k & \hfill n_k\hfill{} \\[2pt] \hline\hline
\str 1 &  6 \\ 
\str 2 &  15 \\ 
\str 3 &  30 \\ 
\str 4 &  114 \\ 
\str 5 &  522 \\ 
\str 6 &  2529 \\ 
\str 7 &  12636 \\ 
\str 8 &  69744 \\ 
\str 9 &  405168 \\ 
\str 10 &  2449773\\[2pt] \hline
\end{tabular}
\hfill
\begin{tabular}{|C|R|}
\hline
\str k & \hfill n_k\hfill{} \\[2pt] \hline\hline 
\str 11 &  15261150 \\ 
\str 12 &  97808574 \\ 
\str 13 &  641284110 \\ 
\str 14 &  4287838548 \\ 
\str 15 &  29153498904 \\ 
\str 16 &  201163103922 \\ 
\str 17 &  1406107987374 \\ 
\str 18 &  9941935540692 \\ 
\str 19 &  71017384630734 \\ 
\str 20 &  511976000663130\\[2pt]  \hline
\end{tabular}
\hfill
\begin{tabular}{|C|R|}
\hline
\str k & \hfill n_k\hfill{} \\[2pt] \hline\hline 
\str 21 &   3721663648494978 \\ 
\str 22 &   27257992426100979 \\ 
\str 23 &   201015705767041110 \\ 
\str 24 &   1491738880927589808 \\ 
\str 25 &   11134231701698352462 \\ 
\str 26 &   83548054037756075397 \\ 
\str 27 &   630009551885176580298 \\ 
\str 28 &   4772408720293409689260 \\ 
\str 29 &   36305219849933449704906 \\ 
\str 30 &   277278864871401475773546 \\[2pt] \hline
\end{tabular}
\caption{Instanton numbers of the $(4,1)$-threefold.}
\end{center}
\end{table}
\newpage
\section{The Special Geometry of the (1,1)-Manifold(s)}
\vskip-10pt
\subsection{The Picard-Fuchs Operator}
The most symmetric $4$-dimensional reflexive polytope is the the
$24$-cell. It admits three order-$24$ group actions that permute the
$24$ vertices simply transitively such that there are no fixed points
on the Calabi-Yau hypersurface, hence defining $3$ different manifolds
with Hodge numbers $h^{1,1}{\,=\,}h^{2,1}{\,=\,}1$. For a full description see~\cite{Braun:2011hd}. In each of these three cases, the Calabi-Yau hypersurface in the covering space toric variety is again of the form
\beqnn
  P_\vph ~=~
  \prod_{i=1}^{24} z_i - \vph \sum_{i=1}^{24} z^{v_i} ~=~ 
  t_1 t_2 t_3 t_4  - \vph \sum_{i=1}^{24} t^{v_i} ~=~
  t_1 t_2 t_3 t_4 \big( 1- \vph f(t) \big),
\eeqnn
written either in homogeneous coordinates $\{z_i\}_{i=1,\dots,24}$ or
in inhomogeneous coordinates $t_1, \dots, t_4$. So we have to again compute $[f^n(t)]_0$. This time, the
Laurent polynomial is
\begin{equation}
  \begin{split}
    f(t) ~&=~
    t_1 +t_2+t_3+t_4+\frac{t_1 t_4}{t_2 t_3}+\frac{t_4}{t_3}+\frac{t_4}{t_2}+
    \frac{t_4}{t_1}+\frac{t_1}{t_4}+\frac{t_2}{t_4}+\frac{t_3}{t_4}+
    \frac{t_2 t_3}{t_1 t_4}+\frac{t_1}{t_2 t_3}\\[5pt]
     &\hspace{53pt}+\frac{1}{t_3}+\frac{1}{t_2}+\frac{1}{t_1}+\frac{t_1}{t_2}+
    \frac{t_1}{t_3}+\frac{t_2 t_3}{t_4}+\frac{t_2 t_3}{t_1}+
    \frac{t_2}{t_1}+\frac{t_3}{t_1}+\frac{t_4}{t_2 t_3}+\frac{1}{t_4}
    \\[10pt]
    ~&=~
    T(t_1,t_2) + T(t_3,t_4) + 
    T\left(\frac{t_1}{t_3},\frac{t_2}{t_4}\right) +
    T\left(\frac{t_1}{t_4}, \frac{t_2 t_3}{t_4}\right),
  \end{split}
\end{equation}
where we used the same expression for $T$ as before in
eq.~\eqref{eq:Tdef}. On this occasion, factors originating from different
$T$-summands can cancel in powers $f^n$, so we need the full
expansion for the powers of $T(t_1,t_2)$
\beqnn
T(t_1, t_2)^k~=~\sum_{\a,\b=-k}^k T_{k\,\a\b}\,t_1^\a t_2^\b~.
\eeqnn
Explicitly, we have
\beqnn
  \begin{split}
    T_{k\,\a\b}~&=\!\! 
    \sum_{\raisebox{-3pt}{$\scriptstyle\r\,=\,\text{max}(\a,\, 0)$}}^{
      \raisebox{4pt}{$\scriptstyle\lfloor\frac{k+\a}{2}\rfloor$}} ~~
    \sum_{j\,=\,\text{max}\left(\lceil\frac{k+2\a+\b-3\r}{2}\rceil,\, 0\right)}^{
      \text{min}\left(k+\a-2\r,\,\lfloor\frac{k+\a+\b-\r}{2}\rfloor\right)} 
   ~\frac{(2\r-\a)!}{\r!\, (\r-\a)!}~\times\\[5pt]
    &\hskip60pt \times
    \frac{k!}{ j!\, (3\r+2j-k-2\a-\b)!\, (k+\a+\b-\r-2j)!\, (k+\a-2\r-j)!}~.
  \end{split}
\eeqnn
The powers of the polynomial are
\beqnn
  f(t)^n = 
  \sum_{\substack{a,b,c,d\,\geq\, 0 \\[2pt] a+b+c+d=n}} 
  \frac{n!}{a!\, b!\,  c!\, d!}
  \,T^a(t_1,t_2)  \,T^b(t_3,t_4)  
  \,T^c\left(\frac{t_1}{t_3},\frac{t_2}{t_4}\right)
  \,T^d\left(\frac{t_1}{t_4},\frac{t_2 t_3}{t_4}\right)
\eeqnn
On substituting the expansions for the $T$'s and picking out the terms
independent of the $t_i$, we obtain the fundamental period as
\beqnn
  \varpi_0(\vph)
  =\sum_{n=0}^\infty \hskip-3pt  \sum_{
    \begin{subarray}{c}
      a,b,c,d \geq 0 \\
      a+b+c+d = n
    \end{subarray}
  } \!\!
  \frac{n!}{a!\, b!\, c!\, d!}\, \vph^n\hskip-5pt
  \sum_{\a,\b\,=-a}^a \hskip3pt\sum_{\g,\d\,=-b}^b \!\!T_{a\,\a\b}\,
  T_{b\,\g\d}\, T_{c,\,\a+\b+\d,\, -(\a+\g+\d)}\,
  T_{d,\,\b+\d,\,-(\a+\b+\g+\d)}\,.
\eeqnn
This expression is somewhat involved but presents little difficulty to machine computation.

The fundamental period again must obey a fourth-order differential
equation of the form \eqref{eq:PFeqn}.
Expanding $\varpi_0$ to sufficiently high order and solving for
polynomials $R_j$ yields degree-$7$ coefficients 
\beqnn\begin{split}
 R_4&= 8957952 \left(\vph + \smallfrac{1}{18}\right)^2
    \left(\vph + \smallfrac13\right)  \left(\vph + \smallfrac14\right)  
    \left(\vph + \smallfrac18\right)  \left(\vph + \smallfrac{1}{12}\right)  
    \left(\vph - \smallfrac{1}{24}\right)   \\[10pt]
 R_3&= 36\vph \left(\vph + \smallfrac{1}{18}\right) \left(1990656 \vph^5+1257984 \vph^4+264384 \vph^3+22320\vph^2 +800 \vph +15\right) \\[15pt]
R_2&= \vph  \big( 206032896 \vph^6+118195200 \vph^5+24103872 \vph^4+2276640 \vph^3
 +105552 \vph^2\\
&\hskip11.12cm+2114 \vph +19 \big) \\[10pt]
R_1&= 72\vph \left(\vph + \smallfrac{1}{18}\right) \left(3483648 \vph^5+1548288 \vph^4+
 225072 \vph^3+13572\vph^2+320 \vph +1\right) \\[15pt]
R_0&= 96 \vph^2 \left(1119744 \vph^5+508032 \vph^4+82512 \vph^3+6318 \vph^2+
 237\vph +4\right)~. \\
\end{split}\eeqnn

As already noted, this Picard-Fuchs equation has been found
previously~\cite{Guttmann:2009, Almkvist:2005} in a non-geometric
context, just from studying the generating function for walks in the
$4$-d face-centered\footnote{That is, $2$-face centered: there is an
  extra lattice point in the center of each $2$-face of the simple
  cubic lattice. Note that, in $4$-d, these automatically generate a
  lattice point in the center $(1,1,1,1)$ of the cubic~cell.}
hyper-cubic (fcc) lattice. We will explore the connection with walks
in detail in \sref{sec:LatticeWalks}. For now note that the naive
generators of this lattice are
\begin{equation}
  \big\{
  (2,0,0,0),
  \dots,\,
  (0,0,0,2),\,
  (1,1,0,0),\,
  (1,0,1,0),\,
  (1,0,0,1)
  \big\},
\end{equation}
and the nearest neighbors of the origin indeed span a polytope that is
$\text{GL}(4,\Q)$-equivalent to the reflexive $24$-cell. However, it has
twice the lattice volume, hence it is not $\text{SL}(4,\Q)$-equivalent. The
fcc nearest neighbor polytope, that is, the polytope whose vertices
are all permutations and sign flips of $(1,1,0,0)$, is not reflexive
and we cannot associate a toric Calabi-Yau hypersurface to
it. However, it is combinatorially the same as the lattice generated
by the vertices of the lattice $24$-cell. In fact, the presentation as
the fcc lattice is easier, and~\cite{Broadhurst:2009, Guttmann:2010nj}
give the nicer generating series
\begin{equation}
\begin{split}
  \varpi_0(\vph) ~&= \hskip-5pt\sum_{i,j,k,l,m=0}^\infty  
  \binom{2i}{i} \binom{2j}{j} \binom{2k}{k} 
  \binom{l+m}{m} \binom{2(l+m)}{l+m}^2\times 
  \\[7pt]
  &\hskip30pt
  \times
  \binom{i+j+k+l+m}{2(l+m)} \binom{i+j+k-l-m}{-i+j+k}
  \binom{2i-l-m}{i-k-l}\, \vph^{i+j+k+l+m}.
\end{split}
\end{equation}

The full set of solutions to the Picard Fuchs equation is as in \eqref{FrobBasis}.
With the series $h_j$ given~by
\begin{equation}\begin{split}
h_0 ~&=~1 + 24 \,\vph^2 + 192 \,\vph^3 + 3384 \,\vph^4 +  51840 \,\vph^5 + 911040 \,\vph^6 + \cO(\vph^7) 
\\[3mm]
h_1 ~&=~ 4 \,\vph + 40 \,\vph^2 + \frac{1816}{3} \,\vph^3 + 8644 \,\vph^4 + \frac{753744}{5} \,\vph^5 + 
\frac{7747372}{3} \,\vph^6 +\cO(\vph^7)
\\[3mm]
h_2 ~&= -6 \,\vph -\frac{51}{2} \,\vph^2 -\frac{3062}{3} \,\vph^3 -\frac{299593}{24} \,\vph^4 
-\frac{19639868}{75} \,\vph^5 - \frac{44274137}{10} \,\vph^6 +\cO(\vph^7)
\\[3mm]
h_3 ~&= -36 \,\vph -\frac{57}{2} \,\vph^2  - \frac{6346}{3} \,\vph^3 - \frac{231785}{16} \,\vph^4
-\frac{71264197}{250} \,\vph^5 - \frac{821688758}{225} \,\vph^6 +\cO(\vph^7)
\end{split}\notag\end{equation}

The singularities of the Picard-Fuchs equation are evident from the
factorization of $R_4$. The behavior of solutions around each of the
singularities is summarized by the Riemann symbol
\begin{equation}\label{eq:RiemannP}
  \renewcommand{\arraystretch}{1.5}
  \mathcal{P}
  \left\{
    \begin{array}{rrrrrrrrc}
      ~0 & 
      ~\infty & 
      -\tfrac{1}{3} &
      -\tfrac{1}{4} &
      -\tfrac{1}{8} &
      -\tfrac{1}{12} &
      -\tfrac{1}{18} &
      ~\tfrac{1}{24} 
      \\
      0 &     
      1 &     
      0 &     
      0 &     
      0 &     
      0 &     
      0 &     
      0 &     
      \\
      0 &     
      2 &     
      1 &     
      1 &     
      1 &     
      1 &     
      1 &     
      1 &     
      \\
      0 &     
      2 &     
      1 &     
      1 &     
      1 &     
      1 &     
      3 &     
      1 &     
      \\
      0 &     
      3 &     
      2 &     
      2 &     
      2 &     
      2 &     
      4 &     
      2 &     
    \end{array}
    ~\vph\right\},
\end{equation}
where we again notice the point of maximal unipotent monodromy at
$\vph=0$ and a point with four power series solutions (and, therefore, no monodromy) at 
$\vph=-\tfrac{1}{18}$. We will see that $\vph{\,=\,}\infty$ corresponds to a conifold point. This suggests a change of variable that brings this conifold to a finite point and so strengthens the similarity to the situation for the (1,4)-manifold. We will return to this after considering the various monodromy matrices.

\subsection{Monodromy}
The geometric meaning of the singularities requires us to look at the details of the
group actions when descending to the $h^{11}=h^{21}=1$ quotient, and is summarized in \tref{tab:sing}.
\begin{table}[htb]
  \renewcommand{\arraystretch}{1.5}
  \centering
  \begin{tabular}{|c||c|c|l|c|l|c|l|}
    \hline
    \multirow{2}{*}{$\vph$} 
    & \multirow{2}{*}{$\tilde\mu(P_\vph)$} 
    & \multicolumn{2}{c|}{$G_1\simeq \text{SL}(2,3)$}
    & \multicolumn{2}{c|}{$G_2\simeq \Z_3\rtimes \Z_8$}
    & \multicolumn{2}{c|}{$G_3\simeq \Z_3 \times Q_8$}
    \\\cline{3-8}
    &&
    $|S|$ & Description &
    $|S|$ & Description &
    $|S|$ & Description
    \\
    \hline\hline
    \multirow{2}{*}{$-\tfrac{1}{3}\+$} & 
    \multirow{2}{*}{$32$} &
    \multirow{2}{*}{$4$} & 
    \multirow{2}{*}{$\Z_3$-orbifold} &
    $1$ & $\Z_3$-orbifold & $1$ & $\Z_3$-orbifold 
    \\[-2mm]
    &&&& $1$ & conifold & $1$ & conifold 
    \\[2pt]\hline
    $-\tfrac{1}{4}\+$ & 
    $24$ &
    $1$ & conifold &
    $1$ & conifold &
    $1$ & conifold 
    \\[2pt]\hline
    $-\tfrac{1}{8}\+$ & 
    $3$ &
    $1$ & $Q_8$-orbifold &
    $1$ & $\Z_8$-orbifold &
    $1$ & $Q_8$-orbifold 
    \\[2pt]\hline
    $-\tfrac{1}{12}\+$ & 
    $24$ &
    $1$ & conifold &
    $1$ & conifold &
    $1$ & conifold 
    \\[2pt]\hline
    $-\tfrac{1}{18}\+$ &
    $0$ &
    \multicolumn{6}{c|}{\dotfill{} smooth, $S=\emptyset$ \dotfill}
    \\\hline
    $\tfrac{1}{24}$ & 
    $1$ & 
    $1$ & $G_1$-orbifold &
    $1$ & $G_2$-orbifold &
    $1$ & $G_3$-orbifold 
    \\[2pt]\hline
    $\infty$ & 
    $12$ & 
    $1$ & $\Z_2$-orbifold &
    $1$ & $\Z_2$-orbifold &
    $1$ & $\Z_2$-orbifold 
    \\[2pt]\hline
    $0$ &
    $\infty$ &
    \multicolumn{6}{c|}{\dotfill{} large complex structure limit, $|S|=\infty$ \dotfill}
    \\[2pt]\hline
  \end{tabular}
  \caption{Singularities in the complex structure moduli space of the
    one-parameter family $X_\vph^{(i)} = \Xt_\vph/G_i$,
    $i\in\{1,2,3\}$. 
    For example, consider the
    singular variety $X_{-1/3}^{(3)}$. Its covering space has $32$
    isolated singular points, each contributing $1$ to the total
    Milnor number $\tilde\mu=32$. 
    Under the $G_3$-action, they form one orbit of length $8$ and one
    orbit of length $24$. Therefore, $X_{-1/3}^{(3)}$ has two singular
    points, one is a $\Z_3$ quotient of a conifold and the other is
    the usual conifold.
  }
  \label{tab:sing}
\end{table}

Owing to the the logarithms the period vector $\vp$ transform as
$\varpi\mapsto m_0\varpi$, under monodromy around $\vph{\,=\,}0$ with the
same monodromy matrix $m_0$ as in eq.~\eqref{eq:MUM}.

Computing the monodromies numerically around the other singularities,
one finds a rational matrix $m_\vph$ for each. Apart from the large
complex structure $\vph=0$ and the point $\vph=-\tfrac{1}{18}$, which
(surprisingly) corresponds to a smooth manifold, these all have a
single logarithmic period. Hence, we expect them to be of conifold
type again and they should be expressible as
\begin{equation}\label{eq:udef}
  m_\vph = \one - \mu_\vph u_\vph u^T_\vph \sigma^{-1}
\end{equation}
where 
\begin{itemize}
\item $\mu_\vph=\mu(P_\vph)$ is the equivalent number of
  conifolds (that is, Milnor number) on the quotient
  $\Xt=X/G$. Since we can identify, at each fixed point orbit
  $G\cdot p$, a normal subgroup $S_p\subset G$ as the stabilizer, the
  Milnor number on the quotient and Milnor number $\tilde\mu$ on the
  covering space are related as
\beqnn
    |S_p|\; \tilde\mu(p) ~=~ \mu(G\cdot p).
\eeqnn
\item $u_\vph$ is the total vanishing cycle of the conifolds, that
  is, the sum of the vanishing cycles.
\item $\sigma$ is the symplectic form in the Frobenius
  basis (not: the basis where the symplectic form is the
  standard $\Sigma$ matrix eq.~\eqref{eq:symplectic}). As before, we
  have
\begin{equation}
    \sigma ~=~\frac{1}{Y_{111}}
    \left(\begin{array}{rrrr}
      0 & 0 &\hphantom{-} 0 & -6 \\
      0 & 0 & 2 & 0  \\
      0 & -2 & 0 & 0 \\
      6 & 0 & 0 & 0
    \end{array}\right)
\notag\end{equation}
  though for this manifold we have $Y_{111} = 4$ so the
  the anti-diagonal entries of $\sigma$ are different from those for the
  $(4,1)$-manifold in eq.~\eqref{eq:symplectic}.
\end{itemize}
There is a slight complication at $\vph=-\tfrac{1}{3}$ where it turns
out that there are always multiple $G$-orbits that, moreover, arrange
into orbits of different size under $G_1$ vs.\ $G_2$ and $G_3$. In the
following, we simply pick the larger $\mu_{-1/3}=12$ leading to the
shorter vector $u_{-1/3}$. With this convention, all relevant data for
the vanishing cycles is listed in \tref{tab:udata}.
\begin{table}[tbp]
  \renewcommand{\arraystretch}{1.6}
  \centering
  \begin{tabular}{|@{\hskip10pt}c@{\hskip30pt}c@{\hskip35pt}c@{\hskip25pt}l@{\hskip20pt}ll|}
    \hline
    $\vph$ & $\tilde\mu(P_\vph)$ & $\mu$ &\hfill $u_\vph$ &\hfill$U_\vph$ & 
    \\[1mm]\hline\hline
    $-\tfrac{1}{3}\+$ & $32$ & $12$ (or $4$) & $(8, 3, \tfrac{7}{6}, \tfrac{3}{4})$ & $(1, -7, 8, 3)$ &
    \\
    $-\tfrac{1}{4}\+$ & $24$ & $1$ & $(24, 11, 5, \tfrac{13}{4})$ & $(4, -28, 24, 11)$ &
    \\
    $-\tfrac{1}{8}\+$ & $3$ & $8$ & $(3, 2, \tfrac{5}{4}, 1)$ & $(1, -6, 3, 2)$ &
    \\
    $-\tfrac{1}{12}\+$ & $24$ & $1$ & $(0, 1, 1, \tfrac{5}{4})$ & $(1, -4, 0, 1)$ &
    \\
    $\tfrac{1}{24}$ & $1$ & $24$ & $(1, 0, \tfrac{1}{12}, 0)$ & $(0, 0, 1, 0)$ &
    \\
    $\infty$ & $12$ & $2$ & $(12, 3, 1, \tfrac{3}{4})$ & $(1, -6, 12, 3)$ &
    \\[2mm]\hline
  \end{tabular}
  \caption{Vanishing cycles at the six conifold points. The coordinates $u_\vph$ are relative to the $\varpi$-basis, and
    $U_\vph$ is relative to the integral symplectic $\Pi$-basis.}
  \label{tab:udata}
\end{table}

Let us return now to the Riemann symbol and make the change of variable
\beq
\vph~=~\frac{\tilde\vph}{1-18\,\tilde\vph}~,~~~\text{so}~~~\tilde\vph~=~\frac{\vph}{1+18\,\vph}~.
\notag\eeq
Using the properties of the Riemann symbol we see that \eqref{eq:RiemannP} is equivalent to the scheme
\begin{equation}\label{eq:RiemannP2}
  \renewcommand{\arraystretch}{1.5}
 \frac{1}{1+18\,\vph}~ \mathcal{P}
  \left\{
    \begin{array}{rrrrrrrrc}
      ~0 & 
      ~\infty & \tfrac{1}{42} & \tfrac{1}{18} & \tfrac{1}{15} & \tfrac{1}{14} & \tfrac{1}{10} &
      ~\tfrac{1}{6} 
      \\
      0 &     1 &     0 &     0 &     0 &     0 &     0 &     0 &     
      \\
      0 &     2 &     1 &     1 &     1 &     1 &     1 &     1 &     
      \\
      0 &     4 &     1 &     1 &     1 &     1 &     1 &     1 &     
      \\  
      0 &     5 &     2 &     2 &     2 &     2 &     2 &     2 &        
    \end{array}
    ~\frac{\vph}{1+18\,\vph}
  \right\}~.
\end{equation}
The change of variables has interchanged the roles of the singularities at $\vph{\,=\,}\infty$ and 
$\vph{\,=\,}\tfrac{1}{18}$ with those at $\tilde\vph{\,=\,}\tfrac{1}{18}$ and 
$\tilde\vph{\,=\,}\infty$. The six conifold points now appear on an equal footing in the symbol.
Based on this new symbol we can choose new period vector
\beq
\widetilde\vp(\tilde\vph)~=~(1+18\,\vph)\, \vp(\vph)
\notag\eeq
This satisfies a slightly simpler Picard-Fuchs equation with coefficient functions that are sixth order polynomials rather than seventh order, for detail see \cite[\SS4.3]{Broadhurst:2009}.

\subsection{Prepotential}\label{sec:prepotential}
Let us quickly review the general structure of the prepotential. For
simplicity, we take the K\"ahler cone to be simplicial and spanned by
$(1,1)$-forms $\omega_i$,
\beqnn
  \Kcone ~=~\Span_{\IR_\geq}\Big\{e_1,~\dots,~e_{h^{11}} \Big\}~.
\eeqnn
Then, in the special coordinates $(z^0,\dots, z^{h^{11}})$, the
prepotential takes the form 
\beqnn
\Fprepotential(z^0,\dots, z^{h^{11}}) ~=~\Fpert + \Fnonpert
\eeqnn
 whose perturbative part equals
\beqnn
  \begin{split}
    \Fpert 
    =&\; -\frac{1}{3!}\sum_{a,b,c=0}^{h^{11}} \frac{Y_{abc} z^a z^b z^c}{z^0}
    \\[8pt]
    =&\;-\frac{1}{3!}\sum_{i,j,k=1}^{h^{11}}\frac{z^i z^j z^k}{z^0}\!
    \int_X\hskip-5pt e_i e_j e_k   
    -\frac{1}{4}\sum_{j,k=1}^{h^{11}} z^j z^k\!
    \int_X\! i^* i_!(e_j)\, e_k  
    \\[10pt]
    &\hspace{5cm}
    +\frac{1}{24}\sum_{k=1}^{h^{11}}z^0 z^k \!
    \int_X \! c_2(X)\, e_k   
    +\frac{1}{2}(z^0)^2\,
    \chi(X)\frac{\zeta(3)}{(2\pi \ii)^3}~,    
  \end{split}
\eeqnn
where $i:X\hookrightarrow\CP_\nabla$ is the embedding map into the ambient toric variety.

Given an embedding as a hypersurface, the natural $(1,1)$-class is the
pull-back of the canonical class of the ambient space, $i^*
(K_\nabla)$. This is always an integral cohomology class, but might be
further divisible in $H^2(X,\Z)$. For example, the pull-back canonical
class of the quintic is divisible by $5$. If we furthermore consider
free quotients of a toric hypersurface, we have both the embedding and
the quotient map
\beqnn
  i: \Xt \hookrightarrow \CP_\nabla,\quad q: \Xt \longrightarrow X\,.
\eeqnn
Now we have a further divisibility problem, as the pull-back of a
primitive generator in $H^2(X,\Z)$ need not be primitive. For example,
the generator $H^2$ on the $\Z_5{\times}\Z_5$-quotient of the quintic
pulls back (via $q^*$) to $5$ times the generator on the quintic.

However, we note that this cannot happen in solely permutation
quotients, since in this case the regular orbits under the group
action are precisely the invariants. In the case at hand, there is a
unique primitive K\"ahler class $\omega\in H^{1,1}(X,\Z)$. Its
pullback under the quotient map $q:\Xt\to X$ is the primitive
$G$-invariant class $\tilde\omega \in H^{1,1}(\Xt,\Z)$, which itself
is the restriction of the K\"ahler class $\omega_\nabla$ of the
ambient toric variety $\CP_\nabla$. That is,
\beqnn
  q^*(\omega) ~=~ \tilde\omega ~=~ i^*(\omega_\nabla)
\eeqnn
The numerical invariants on the covering space are, of course, simple
computations in toric geometry. For the case at hand, including the
rescaling due to the $G$-quotient, we compute
\beqnn
  \begin{split}
    \int_X e^3 ~=&\;\frac{1}{24} \int_\Xt \tilde{e}^3 ~=~ 4
    \\[8pt]
    \int_X q_! i^*i_!(q^*e)\, e ~=&\;
    \int_\Xt i^*i_!(\tilde e)\, \tilde e ~=~
    \frac{1}{24} \int_{\IP_\nabla} e_\nabla^4 ~=~4
    \\[8pt]
    \int_X c_2(X)\, e ~=&~\frac{1}{24} \int_\Xt c_2(\Xt)\, \tilde{e}~=~4
    \\[10pt]
    \chi(X) ~=&\; \frac{1}{24}\,\chi(\Xt) ~=~ 0~.
  \end{split}
\eeqnn

\subsection{Integral Homology Basis}
Using the numerical invariants above, we obtain
\begin{equation}
  \Fpert(z^0,z^1)~= -\frac{2}{3}\, \frac{(z^1)^3}{z^0} - (z^1)^2 + \frac{1}{6}\, z^0 z^1 ~.
\notag\end{equation}
Is terms of the coefficient $\vph$ of the polynomial, the special
coordinates can be taken to be $(z^0,z^1) = (\varpi_0(\vph), \varpi_1(\vph))$. To relate the period vectors, we look at the asymptotic behavior around the large complex structure limit $t=0$, where
\begin{equation}\label{eq:mirrormap}
  t ~=~ \frac{z^1}{z^0} ~=~\frac{1}{2\pi\ii} \log\vph + \dots
\notag\end{equation}
In terms of the special coordinates, the period vector takes the form
\begin{equation}
  \Pi(z^0,z^1) ~=~ 
  \begin{pmatrix}
    \partial\Fprepotential\big/\partial z^0  \\[3pt]
    \partial\Fprepotential\big/\partial z^1  \\[2pt]
    z^0 \\
    z^1
  \end{pmatrix}
  \sim
  \begin{pmatrix}
    \tfrac{2}{3} t^3 +\tfrac{1}{6} t \\[4pt]
    -2 t^2 - 2t +\tfrac{1}{6} \\[2pt]
    1 \\
    t
  \end{pmatrix}
 ;\qquad\varpi ~=~
  \begin{pmatrix}
    \varpi_0 \\
    \varpi_1 \\
    \varpi_2 \\
    \varpi_3 \\
  \end{pmatrix}
  \sim
  \begin{pmatrix}
    1 \\
    t \\
    t^2 \\
    t^3
  \end{pmatrix}
\notag\end{equation}
Hence, the basis transformation $\Pi = \rho\,\varpi$ between the two
different bases for the periods must be
\begin{equation}\label{eq:integralbasis}
  \rho ~=~ 
  \left(\begin{array}{rrrr}
    0 & \tfrac{1}{6} & 0 & \hphantom{-}\tfrac{2}{3} \\[2pt]
    \tfrac{1}{6} & -2 & -2 & 0 \\[2pt]
    1 & 0 & 0 & 0 \\[2pt]
    0 & 1 & 0 & 0 
  \end{array}\right)
\end{equation}
Expressed in the $\Pi$-basis, the monodromy matrices are $M_\vph =
\rho\, m_\vph \rho^{-1}$. We find
\begin{table}[H]
\begin{equation}
\newcolumntype{P}{>{\raggedleft\arraybackslash}p{0.7cm}}
  \begin{aligned}
    M_{-1/3} - \one~=&~ 
    \left(\hskip-5pt
      \begin{smallarray}{PPPP}
        -96 & -36 & 12 & -84 \\[2pt]
        672 & 252 & -84 & 588 \\[2pt]
        -768 & -288 & 96 & -672 \\[2pt]
        -288 & -108 & 36 & -252
      \end{smallarray}
    \right)
    ,
    &
    M_{-1/4} - \one~=&~ 
    \left(\hskip-5pt
      \begin{smallarray}{PPPP}
        -96 & -44 & 16 & -112 \\[2pt]
        672 & 308 & -112 & 784 \\[2pt]
        -576 & -264 & 96 & -672 \\[2pt]
        -264 & -121 & 44 & -308
      \end{smallarray}
    \right)
    ,
    \\[10pt]
    M_{-1/8} - \one~=&~
    \left(\hskip-5pt
      \begin{smallarray}{PPPP}
        -24 & -16 & 8 & -48 \\[2pt]
        144 & 96 & -48 & 288 \\[2pt]
        -72 & -48 & 24 & -144 \\[2pt]
        -48 & -32 & 16 & -96
      \end{smallarray}
    \right)
    ,
    &
    M_{-1/12}- \one ~=&~
    \left(\hskip-5pt
      \begin{smallarray}{PPPP}
        0 & -1 & 1 & -4 \\[2pt]
        0 & 4 & -4 & 16 \\[2pt]
        0 & 0 & 0 & 0 \\[2pt]
        0 & -1 & 1 & -4
      \end{smallarray}
    \right)
    ,
    \\[10pt]
    M_{-1/18} - \one~=&~
    \left(\hskip-5pt
      \begin{smallarray}{PPPP}
        0 & 0 & 0 & 0 \\[2pt]
        0 & 0 & 0 & 0 \\[2pt]
        0 & 0 & 0 & 0 \\[2pt]
        0 & 0 & 0 & 0
      \end{smallarray}
    \right)
    ,
    &
    M_0 - \one~=&~
    \left(\hskip-5pt
      \begin{smallarray}{PPPP}
        0 & -1 & 1 & 0 \\[2pt]
        0 & 0 & -4 & -4 \\[2pt]
        0 & 0 & 0 & 0 \\[2pt]
        0 & 0 & 1 & 0
      \end{smallarray}
    \right)
    ,
    \\[10pt]
    M_{1/24}  - \one~=&~
    \left(\hskip-5pt
      \begin{smallarray}{PPPP}
        0 & 0 & 0 & 0 \\[2pt]
        0 & 0 & 0 & 0 \\[2pt]
        -24 & 0 & 0 & 0 \\[2pt]
        0 & 0 & 0 & 0
      \end{smallarray}
    \right)
    ,
    &
    M_{\infty} - \one ~=&~ 
    \left(\hskip-5pt
      \begin{smallarray}{PPPP}
        -24 & -6 & 2 & -12 \\[2pt]
        144 & 36 & -12 & 72 \\[2pt]
        -288 & -72 & 24 & -144 \\[2pt]
        -72 & -18 & 6 & -36
      \end{smallarray}    
    \right)
    .
  \end{aligned}
\notag\end{equation}
\caption{The integral monodromy matrices for the singularities of the $(1,1)$-manifold.}
\end{table}

As expected, the monodromy matrices in the $\Pi$-basis are integral,
$M_\vph\in \text{GL}(4, \Z)$ and symplectic, $M_\vph\Sigma M_\vph^T =
\Sigma$. Here, $\Sigma$ is the standard symplectic matrix as in
\eqref{eq:symplectic}.  We conclude that the $M_\vph$ are the
monodromies in an integral symplectic basis for $H^3(X,\Z)$. The
values of $\vph$ where the manifold develops a conifold singularity
still have the special form eq.~\eqref{eq:udef} of the monodromy
matrix. In the integral basis, this is
\begin{equation}
  M_\vph = \one - \mu_\vph U_\vph U^T_\vph \Sigma^{-1}  
  ,\qquad
  U_\vph =\; \rho\, u_\vph,
\notag\end{equation}
and we list them in \tref{tab:udata}. The vectors $U_\vph$ span the
entire $H^3(X,\Z)$ lattice. We note that, as in the three-generation
manifold, the singularity closest\footnote{That is, the singularity
  that determines the radius of convergence of the power series for the
  fundamental period.} to the origin also has a particularly simple
vanishing cycle $U=(0,0,1,0)$ in the integral basis.

\subsection{Instanton Numbers}
\vskip-10pt
The properly normalized Yukawa coupling is
\begin{equation}
  \begin{split}
    y_{\vph\vph\vph} ~&=~
    \frac{1}{\vph^3}\,\exp\left( -\frac{1}{2}\int d\vph\,\frac{R_3(\vph)}{R_4(\vph)}\right)
    \\[8pt]
    ~&=
    -384\; 
    \frac{\vph + \smallfrac{1}{18}}{
      \vph^3
      \left(\vph + \smallfrac13\right) \left(\vph + \smallfrac14\right) 
      \left(\vph + \smallfrac18\right) 
      \left(\vph + \smallfrac{1}{12}\right) \left(\vph - \smallfrac{1}{24}\right)
    }
   \end{split}
\notag\end{equation}
where we have fixed the integration constant to match the classical
intersection number (times $\vph^{-3}$) at the large complex
structure limit $\vph=0$. The mirror Yukawa coupling is
\begin{equation}
  y_{ttt} = 
  \frac{1}{\varpi_0^2} 
  \,y_{\vph\vph\vph}
  \left( \frac{1}{2\pi i} \frac{d\vph}{dt} \right)^3~.
\notag\end{equation}
We have to invert the mirror map eq.~\eqref{eq:mirrormap} to
express $\vph$ in terms of $t$ or, rather, $q \eqdef \exp(2\pi i
t)$. The mirror map and its inverse are
\begin{equation}
  \begin{split}
    q(\vph) ~&=~ \vph + 4 \vph^{2} + 48 \vph^{3} + 680
    \vph^{4} + 10084 \vph^{5} + 173056 \vph^{6} + 3011300
    \vph^{7} + \cO\left(\vph^{8}\right)
    \\[5pt]
    \vph(q)~&=~ q -4 q^{2} -16 q^{3} -40 q^{4}
    + 604 q^{5} -9872 q^{6} + 56092 q^{7} +
    \cO\left(q^{8}\right).
  \end{split}
\notag\end{equation}
We therefore obtain the instanton-corrected Yukawa coupling
\begin{equation}
  \begin{split}
    y_{ttt} \;=&~ 4 + 12 q -116 q^{2} + 6924 q^{3} -64884 q^{4} + 
    2146012 q^{5} -22118516 q^{6} + O\left(q^{7}\right)
    \\[5pt]
    =&~ 4 + \sum_{k=1}^\infty \frac{n_k k^3 q^k }{1- q^k}
  \end{split}
\notag\end{equation}
where the instanton numbers $n_k$  are recorded in \tref{tab:instanton}. Up to an
overall factor of $4$, these have already appeared in~\cite{Broadhurst:2009}
from a purely combinatorial point of view. The extra factor is of
course there to due to the classical intersection number $4$, and
appears to be only meaningful in the geometric setting.
\begin{table}[H]
\begin{center}
\def\str{\vrule height14pt depth5pt width0pt}
\newcolumntype{C}{>{~$}c<{$~}}
\newcolumntype{R}{>{~$}r<{$~}}
\begin{tabular}{|C|R|}
\hline
\str k & \hfill n_k\hfill{} \\[2pt] \hline\hline
\str 1 & 12 \\
\str 2 & -16 \\
\str 3 & 256 \\
\str 4 & -1012 \\
\str 5 & 17168 \\
\str 6 & -102432 \\[2pt]  \hline
\end{tabular}
\hskip10pt
\begin{tabular}{|C|R|}
\hline
\str k & \hfill n_k\hfill{} \\[2pt] \hline\hline 
\str 7 & 1768032 \\
\str 8 & -12810048 \\
\str 9 & 226260008 \\
\str 10 & -1831410544 \\
\str 11 & 33000429000 \\
\str 12 & -286340050052 \\[2pt]  \hline
\end{tabular}
\hskip10pt
\begin{tabular}{|C|R|}
\hline
\str k & \hfill n_k\hfill{} \\[2pt] \hline\hline 
\str 13 & 5252822116016 \\
\str 14 & -47718467477584 \\
\str 15 & 890108488876160 \\
\str 16 & -8340130846927456 \\
\str 17 & 158096635640838140 \\
\str 18 & -1512328959263997360 \\[2pt]  \hline
\end{tabular}
\caption{Instanton numbers of the $(1,1)$-threefold.}
\label{tab:instanton}
\end{center}
\vskip-15pt
\end{table}

\subsection{Cross Ratios}
\vskip-10pt
The cross ratio 
\begin{equation}
  X(z_1,z_1; z_2, z_3) = 
  \frac{(z_1-z_3)(z_2-z_4)}{(z_2-z_3)(z_1-z_4)}
\notag\end{equation}
is invariant under reparametrizations of the $\CP^1$.
Some of the cross-ratios of the complex structure singularities for
the $(1,1)$ and $(4,1)$ manifold are the same:
\begin{equation}
  X\left(
    0, -\tfrac{1}{4}, -\tfrac{1}{5}, -\tfrac{1}{6}
  \right)
  = 
  X\left(
    0, -\tfrac{1}{4}, -\tfrac{1}{8}, -\tfrac{1}{12}
  \right)
\notag\end{equation}
But not all cross ratios agree, proving that the complex structure
moduli spaces are different.

One might wonder whether there is any remaining symmetry of the
1-dimensional complex structure parameter spaces. However, one can
check that there is no permutation of the $8$ singularities preserving
the cross ratios. In other words, there is no non-trivial subgroup of
$\text{PSL}(2,\C)$ mapping the singularities to themselves. Hence, the
complex structure moduli space has no holomorphic\footnote{The only
symmetry is complex conjugation.} symmetries.
\begin{figure}[p]
  \centering
  \includegraphics[width=\linewidth]{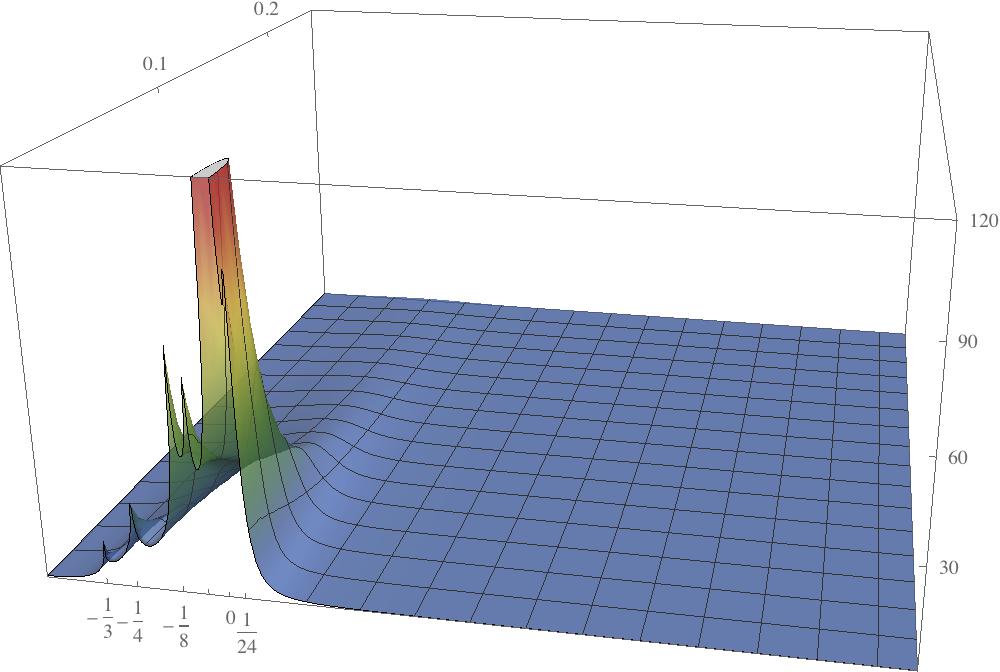}\\
  \vspace{0.5cm}
  \includegraphics[width=0.95\linewidth]{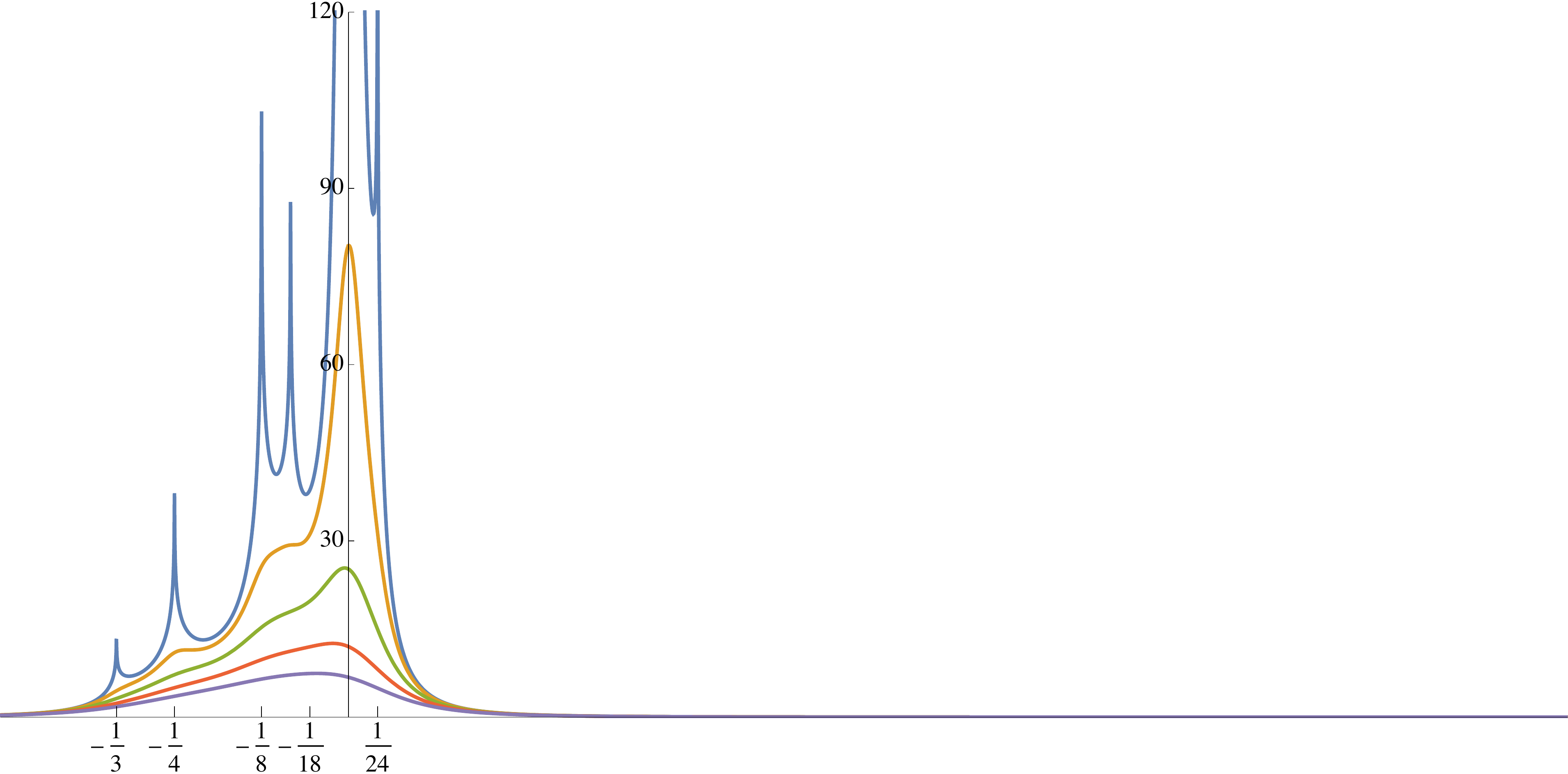}\\
  \vspace{5pt}
  \caption{The metric $g_{\vph\bar\vph}$ on the complex structure moduli space. Top: plot over a region of the complex plane. Bottom: restriction to a real interval and its shift to constant imaginary values. These
figures are included for comparison with \fref{fig:metric} and are plotted to cover the same range of $\vph$.}
\label{fig:oneonemetric}
\end{figure}
\newpage
\section{Lattice Walks}\label{sec:LatticeWalks}
\vskip-10pt
\subsection{Spectral Considerations}
\vskip-10pt
The countably infinite vertices of the lattice can be enumerated, say,
$\{v_i\}_{i\in\Z}$. The allowed steps of the walk then define a matrix
\vskip-10pt
\begin{equation}
  A_{ij} ~=~ 
  \begin{cases}
    1 & \text{if we can step from $v_i$ to $v_j$}, \\[3pt]
    0 & \text{else.}
  \end{cases}
\notag\end{equation}
This infinite symmetric matrix is the adjacency matrix of the graph
whose nodes are the vertices and the edges are the admissible
steps. By it being symmetric we can apply the spectral theorem and
write
\vskip-20pt
\begin{equation}
  A ~=~ \int \lambda \cdot \rho_\lambda \rho_\lambda^T d\lambda,
\notag\end{equation}
where the integral ranges over all eigenvalues with associated
(orthonormal) eigenvector density $\rho_\lambda$. Of particular
interest will be the diagonal entries. Since there is no preferred
vertex, the measure part must all be the same and equal to the
eigenvalue density $\mu(\lambda)$:
\begin{equation}
  |\rho_{\lambda,i}|^2 d\lambda ~=~ \mu(\lambda) d\lambda
  \quad \forall i\;.
\notag\end{equation}
Due to the combinatorics of the matrix product, the powers
$(A^n)_{ij}$ count the number of lattice walks from $v_i$ to
$v_j$. Hence, the generating function for the number of lattice walks
of length $n$ starting and ending at the origin is
\begin{equation}
  \label{eq:spec}
  \varpi_0(\varphi) ~=~ 
  \sum_n (A^n)_{0,0} \,\varphi^n ~=~
  \sum_n \int (\varphi \lambda)^n 
  \mu(\lambda) d\lambda ~=~
  \int 
  \frac{
    \mu(\lambda) d\lambda
  }{
    1 - \varphi \lambda
  }
\notag\end{equation}
We note that the power series must diverge outside of a finite disc
around the origin, but right hand side need not. Of course this is
nothing but the Borel resummation of the series. 

The lattice Laplacian is the closely related matrix $\Delta = D - A$
where $D$ is the \emph{degree} matrix, that is, the diagonal matrix
whose entry $D_{ii}$ equals the number of adjacent nodes of the $i$-th
node. In particular, in all cases that we will consider, the degree
matrix is the identity matrix times the number of nearest neighbors
$a$ of the lattice. As in the continuous case, the spectrum of the
Laplacian is real and non-negative. It is also bounded above by the
lattice cutoff $\Lambda$, hence a subset of a finite real
interval. Therefore, the spectrum of the adjacency matrix is also
contained in a finite real interval
\begin{equation}
  \mathop\mathrm{Spec} \Delta = [0, \Lambda]
  \quad\Leftrightarrow\quad
  \mathop\mathrm{Spec} A = [a-\Lambda, a]
\notag\end{equation}
The integrand of eq.~\eqref{eq:spec} is only well-defined away from
the reciprocal eigenvalues, that is, on
$(-\infty,\tfrac{1}{a-\Lambda}] \cup [\tfrac{1}{\Lambda}, \infty)$. It
is also bounded by $1/(1-\varphi \hat\lambda)$, where
$\hat\lambda = \max(|a-\Lambda|, a)$ is the largest absolute value
eigenvalue of $A$. Hence it converges on the disc $|\varphi|\leq 1/\hat\lambda$, which therefore is the disk where the
power series converges. The power series expression for
$\varpi_0(\varphi)$ necessarily diverges outside of that disk, but
eq.~\eqref{eq:spec} need not. In fact, away from the real line the
denominator is uniformly bounded
\begin{equation}
  \left| 1-\varphi \lambda \right| ~=~ \sqrt{\big(1-\lambda \re(\varphi) \big)^2 + \big(\lambda \im(\varphi) \big)^2} 
  ~\geq~  
  \begin{cases}
    \frac{1}{4}\,; & 
    \text{if}~ 
    \left|\lambda \re(\varphi)\right| \leq \frac{1}{2} \\[8pt]
    \frac12\left|\frac{\im(\varphi)}{\re(\varphi)}\right|\,;
    & \text{else},
  \end{cases}
\notag\end{equation}
hence eq.~\eqref{eq:spec} converges everywhere away from the real
line. We conclude that the singularities of the generating function
(or fundamental period) $\varpi_0(\varphi)$ are restricted to the real
line, and start at distance $1/\hat\lambda$ away from the origin.

\subsection{Square Lattice}
\vskip-10pt
As an example, consider the square/cubical/hyper-cubical
$d$-dimensional lattice $\Z^d$ and allow steps to the $2d$ adjacent
vertices. Equivalently, consider the Calabi-Yau hypersurface in
$\big(\IP^1\big)^d$. Let $c_n^{(d)}$ be the number of lattice walks
starting and ending at the origin. The sum of coordinates mod $2$
always changes at each step, so there can only be walks of even
length: $c_{2n+1}^{(d)} = 0$. In low dimensions, explicit formula are
known to be
\vskip-10pt
\begin{equation}
  c_{2n}^{(1)} ~=~ \binom{2n}{n}
  ,\quad
  c_{2n}^{(2)} ~=~ \binom{2n}{n}^2
\notag\end{equation}
though the apparent pattern does not continue with the cubic (3d)
lattice. The generating function of the square lattice can be
expressed in terms of the complete elliptic integral of the first
kind\footnote{Note that $K(x)=\text{\texttt{elliptic\_kc(x\^{}2)}}$ in Sage.}
as
\vskip-20pt
\begin{equation}
  \varpi_0^{(2)}(\varphi) ~=~ 
  \sum c_{2n}^{(2)} \varphi^{2n} ~=~
  \sum \binom{2n}{n}^2 \varphi^{2n} ~=~
  \frac{2}{\pi} K(4\varphi),
\notag\end{equation}
which is singular at $\varphi = \pm \frac{1}{4}$. We will follow the
conventional choice of branch cuts $(-\infty, -\frac{1}{4}] \cup
[\frac{1}{4}, \infty)$. The power series of course diverges outside of
the disc $|\varphi|\leq\frac{1}{4}$.

\begin{figure}[H]
  \centering
  \includegraphics[width=0.8\linewidth]{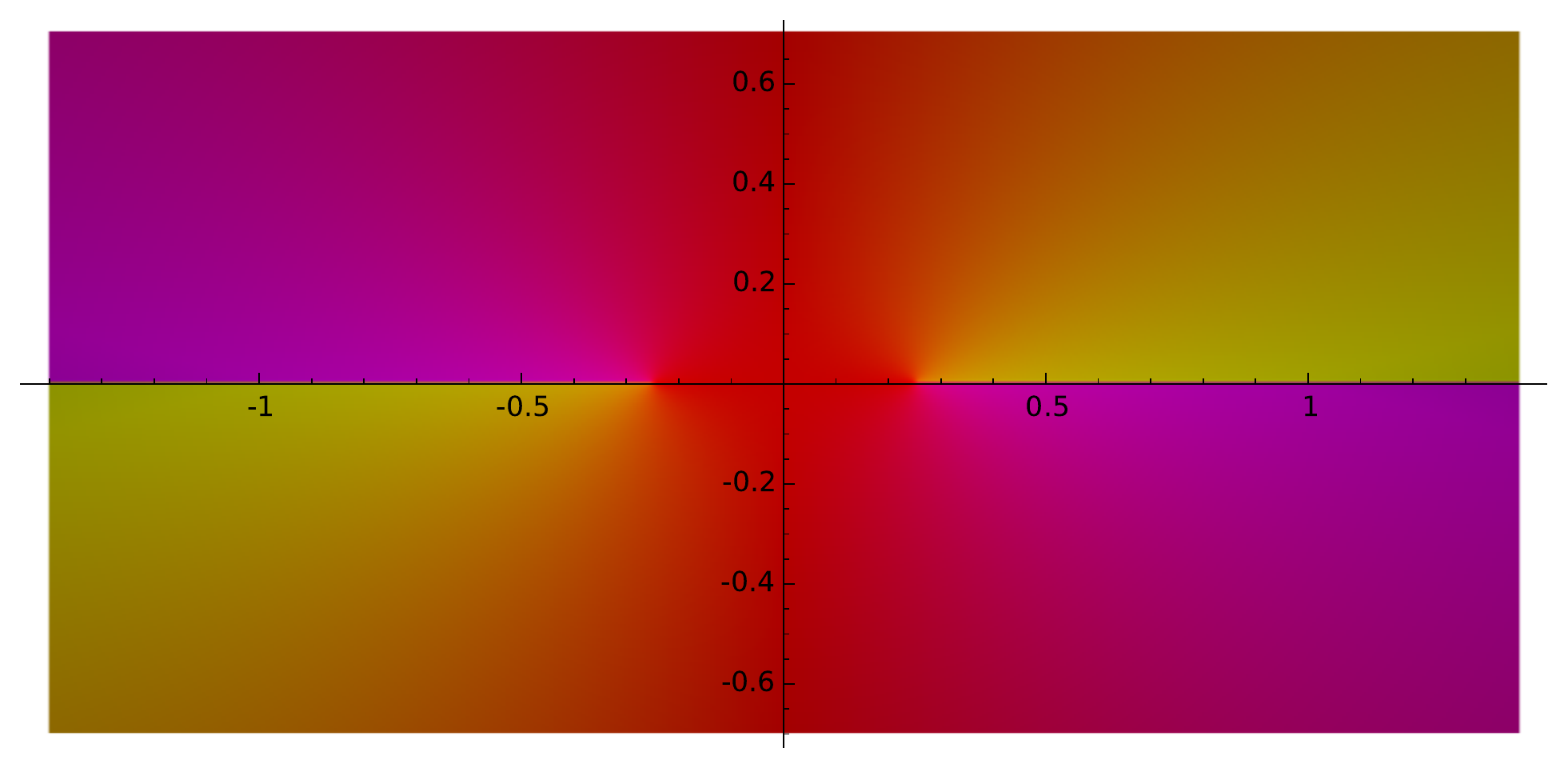}
  \\
  \includegraphics[width=0.8\linewidth]{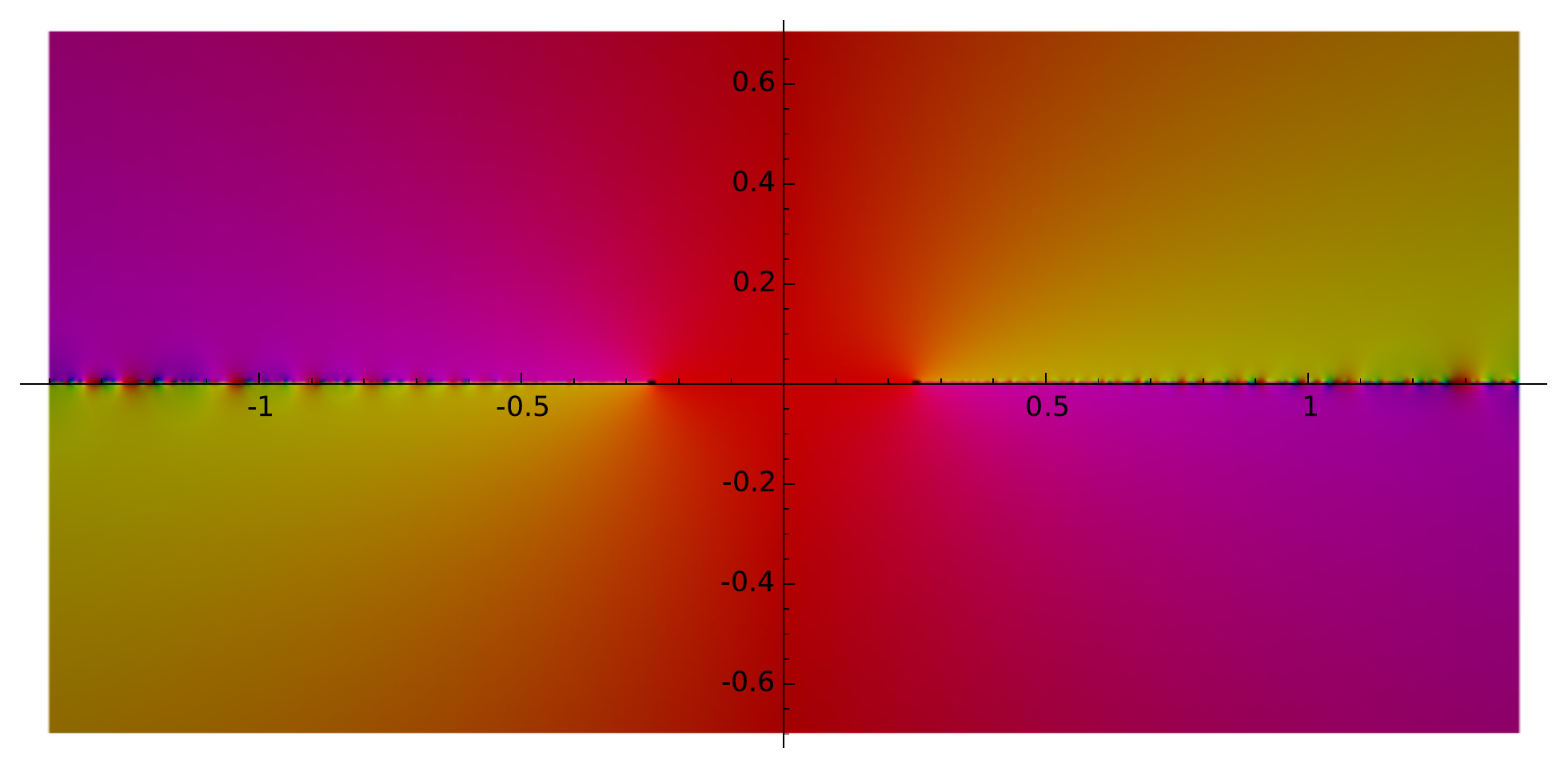}
  \\
  \includegraphics[width=0.8\linewidth]{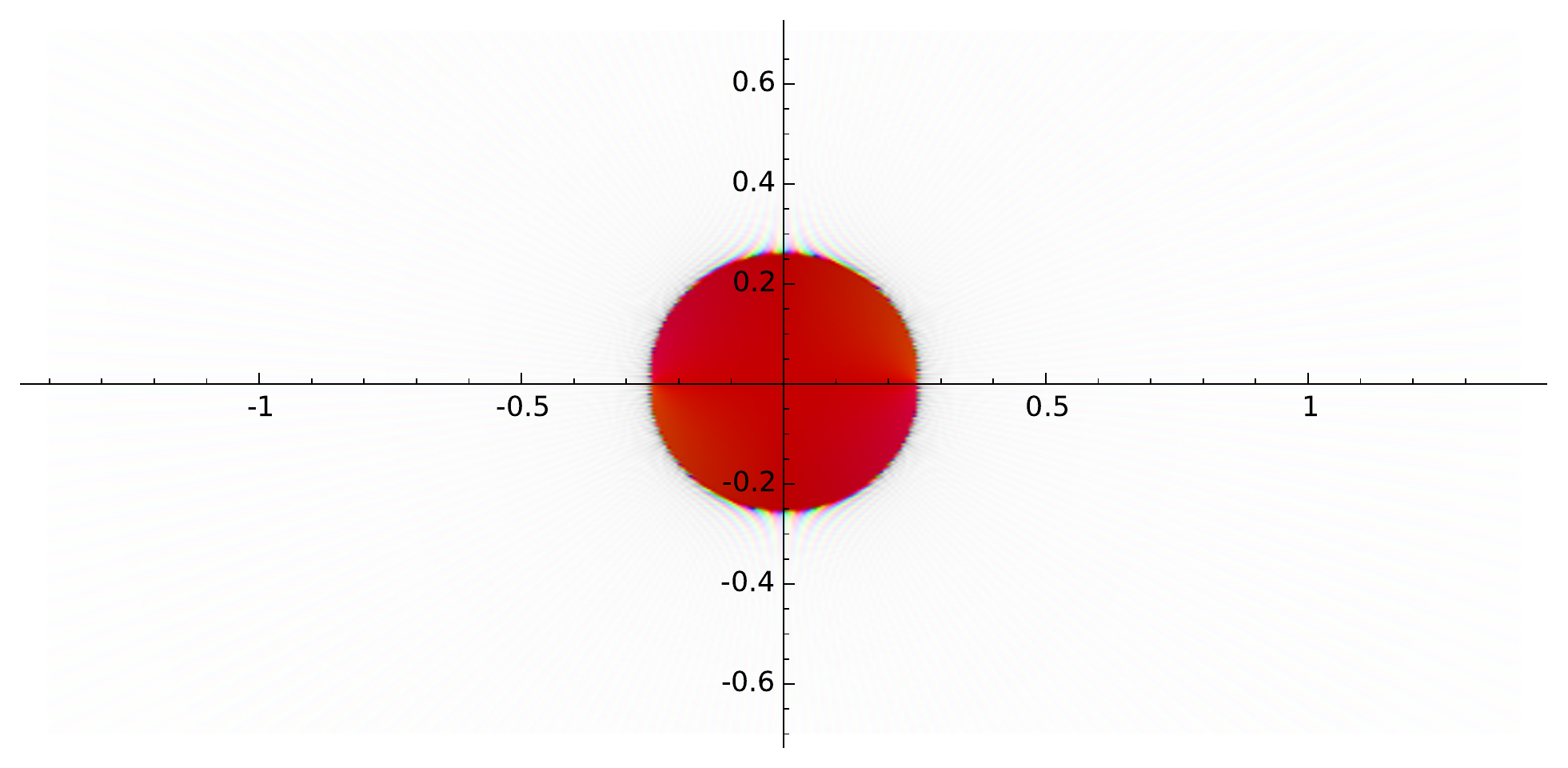}
  \caption{The square lattice generating function. Top: The exact
    solution using the complete elliptic integral. Middle: The spectral sum eq.~\eqref{eq:spectralsum} with $n=75$. 
    Bottom: The truncated series $ \sum_0^{99} c_{2n}^{(2)} \varphi^{2n}$. Brightness is magnitude and hue is phase.}
  \label{fig:square_lattice}
\end{figure}

There are various finite-dimensional approximations to the infinite
lattice, for the purposes of this example we consider it on a torus
with $n\geq 3$ lattice points\footnote{We assume that $n\geq 3$ to
  ensure that the number of adjacent vertices is still $2d$.} in each
direction, for a total of $n^d$ lattice points. For $d=1$, we get the
circle graph whose adjacency matrix has eigenvalues
\begin{equation}
  A^{(1)} = 
  \left(
    \begin{smallmatrix}
      0 & 1 & 0 & 0 & $\normalsize\ldots$ & 0 & 1 \\
      1 & 0 & 1 & 0 & $\normalsize\ldots$ & 0 & 0 \\
      0 & 1 & 0 & 1 & $\normalsize\ldots$ & 0 & 0 \\
      0 & 0 & 1 & 0 & $\normalsize\ldots$ & 0 & 0 \\[-4pt]
      \vdots & \vdots & \vdots & \vdots & \ddots & \vdots & \vdots \\[2pt]
      0 & 0 & 0 & 0 & $\normalsize\ldots$ & 0 & 1 \\
      1 & 0 & 0 & 0 & $\normalsize\ldots$ & 1 & 0
    \end{smallmatrix}
  \right)
  \quad \Rightarrow \quad
  \lambda^{(1)}_i \,=~ 
  2 - 4 \sin\left(\frac{\pi i}{n}\right)^2
  ,\quad
  i = 0, \dots, n-1.
\notag\end{equation}
The eigenvalues for the Cartesian product graph are the sums of the
factor eigenvalues, that~is,
\begin{equation}
  \lambda^{(d)}_{(i_1, \dots, i_n)} ~=~ 
  2d - 4 \sum_{j=0}^d \sin\left(\frac{\pi i_j}{n}\right)^2
  ,\quad
  (i_1, \dots, i_d) \in \{ 0, \dots, n-1 \}^d.
\notag\end{equation}
The spectral version of the square lattice generating function is,
therefore,
\begin{equation}
  \label{eq:spectralsum}
  \varpi_0^{(2)}(\varphi) ~=~ 
  \lim_{n\to \infty}
  \frac{1}{n^2}
  \sum_{i, j=0}^{n-1}
  \frac{
    1
  }{
    1 - 4 \varphi\, \Big(1-\sin(\tfrac{\pi i}{n})^2 - \sin(\tfrac{\pi j}{n})^2 \Big)
  } 
\notag\end{equation}
By truncating the power series and spectral series presentations, we
can compute approximations. In \fref{fig:square_lattice}, we plot
all three and visualize their convergence.
\newpage
\section*{Acknowledgments}
\vskip-10pt
The research of each of the three authors is supported by the EPSRC grant BKRWDM00.
\vskip20pt
\section*{Note Added}
\vskip-10pt
After this paper was completed we became aware of the paper~\cite{Kawada:2015vwe} by Kawada {\em et al.\/}
which has a substantial overlap with the present work.
\vskip20pt
\renewcommand{\baselinestretch}{1.05}\normalsize
\bibliographystyle{utcaps}
\bibliography{OneParameter}
\end{document}